\documentclass[amssymb,amsmath,prd,twocolumn,superscriptaddress,nofootinbib,floatfix]{revtex4-2}

\usepackage{booktabs}
\usepackage{graphicx}
\usepackage{bm}
\usepackage{amssymb,amsmath}
\usepackage{mathrsfs}
\usepackage{latexsym}
\usepackage{color}
\usepackage{placeins}
\usepackage[normalem]{ulem}
\usepackage{dcolumn}
\usepackage[colorlinks=true,citecolor=blue,urlcolor=blue]{hyperref}
\usepackage[usenames,dvipsnames]{xcolor}
\usepackage{xspace}
\usepackage{multirow}
\usepackage{soul}
\usepackage{orcidlink}
\graphicspath{{./}{figures/}}

\allowdisplaybreaks

\definecolor{kcmagenta}{rgb}{0.54, 0.17, 0.88}
\definecolor{kkbrown}{rgb}{0.82,0.41,0.12}
\definecolor{smgreen}{rgb}{0.26, 0.625, 0.277}
\definecolor{dfblue}{rgb}{0.15, 0.75, 0.79}
\definecolor{recomment}{rgb}{0.85, 0.1, 0.1}

\newcommand{\AC}{\affiliation{Department of Physics and Astronomy, Amherst College, Amherst, MA 01002, USA}}
\newcommand{\CIT}{\affiliation{Department of Physics, California Institute of Technology, Pasadena, California 91125, USA}}
\newcommand{\CITLab}{\affiliation{LIGO Laboratory, California Institute of Technology, Pasadena, California 91125, USA}}
\newcommand{\UT}{\affiliation{Center for Gravitational Physics and Department of Physics, The University of Texas at Austin, Austin, TX 78712}}
\newcommand{\UIUC}{\affiliation{Department of Physics, University of Illinois Urbana-Champaign, Illinois 61801, USA}}

\newcommand{\chieff}{\chi_{\textrm{eff}}}
\newcommand{\chip}{\chi_\mathrm{p}}
\newcommand{\Msun}{M_\odot}

\begin{document}

\title{Mapping Parameter Correlations in Spinning Binary Black Hole Mergers}

\author{Karen Kang~\orcidlink{0009-0001-9152-1359}}
\email{kkang25@amherst.edu}
\AC

\author{Simona J.~Miller~\orcidlink{0000-0001-5670-7046}}
\email{smiller@caltech.edu}
\CIT \CITLab

\author{Katerina Chatziioannou~\orcidlink{0000-0002-5833-413X}} 
\email{kchatziioannou@caltech.edu}
\CIT \CITLab 

\author{Deborah Ferguson~\orcidlink{0000-0002-4406-591X}} 
\email{dferg@illinois.edu}
\UT \UIUC 

\date{\today}

\begin{abstract}
 The spins of binary black holes measured with gravitational waves provide insights about the formation, evolution, and dynamics of these systems. However, interpreting these measurements—especially for heavy black holes—remains an open problem. While the imprint of spin during the inspiral phase, where the black holes are well-separated, is understood through analytic descriptions of the dynamics, no such expressions exist for the merger.
 Though numerical relativity simulations provide an exact solution (to within numerical error), the imprint of the full six spin degrees of freedom on the signal is not transparent.
 In the absence of analytic expressions for the merger and to advance our ability to interpret massive binary black hole spin measurements, here we propose a waveform-based approach. 
 Leveraging a neural network to efficiently calculate mismatches between waveforms, we identify regions in the parameter space of spins and mass ratio that result in low mismatches and thus similar waveforms. 
 We map these regions with a Gaussian fit, thus identifying correlations and quantifying their strength.
 For low-mass, inspiral-dominated systems, we recover the known physical imprint: larger aligned spins are correlated with more equal masses as they have opposite effects on the inspiral length.
 For high-mass, merger-dominated signals, a qualitatively similar correlation is present, though its shape is altered and strength decreases with larger total mass. 
 Correlations between in-plane spins and mass ratio follow a similar trend, with their shape and strength altered as the mass increases. Our new methodology of waveform-based correlation mapping provides a first step toward systematically modeling spin effects in merger-dominated signals across the full intrinsic parameter-space and motivates future effective spin parameters beyond the reach of analytic methods.
\end{abstract}

\maketitle

\section{Introduction}
\label{sec:intro}

Quasicircular black hole binaries (BBHs) observed through gravitational waves (GWs) by LIGO~\cite{aLIGO}, Virgo~\cite{aVirgo}, and KAGRA~\cite{KAGRA} are characterized by the component masses and spins.
Though less well measured than the masses, the magnitudes and directions of the BH spins carry imprints of the evolutionary history of the binary and its constituents, e.g., Ref.~\cite{Mandel:2018hfr}.   
Measurements of BH spins from both individual events and the population as a whole are therefore central to the astrophysical interpretation of the observed signals, e.g., Refs.~\cite{LIGOScientific:2020kqk,KAGRA:2021duu}.

Inference of BBH parameters hinges on waveform models for the GW signal as observed in the detectors~\cite{LIGOScientific:2019hgc,GWTC2.1,GWTC2,GWTC3}.
During the inspiral phase, these models are based on analytic equations that are valid in the regime of low-orbital-velocity (compared to the speed of light): the post-Newtonian (PN) expansion~\cite{Blanchet:2013haa}.
The subsequent merger phase is not analytically tractable in any perturbative scheme.
Instead, waveform models are based on numerical relativity (NR) simulations of the binary's full dynamics either by calibrating phenomenological terms, e.g., Refs.~\cite{Pratten:2020ceb,Ramos-Buades:2023ehm} or by directly interpolating the simulations~\cite{Varma:2019csw}.

A general analytic understanding of a problem helps identify the most appropriate parameters for studying its behavior.
For example, the first term in the PN expansion of the inspiral GW phase introduces a certain mass combination,  the chirp mass~\cite{Blanchet:2013haa,Buonanno:2009}.
On the spin front, constraints are expressed through the effective aligned and precessing spins, both again motivated by PN equations.
The effective aligned spin $\chieff$ is related (but not exactly equal) to the leading-order 1.5PN spin-orbit term in the GW phase~\cite{Ajith:2009bn,Arun:2009}. 
It characterizes the spin components in the direction of the Newtonian orbital angular momentum and is conserved under precession to at least the 2PN order~\cite{Racine:2008qv}.
The effective precessing spin $\chip$ is instead motivated by the 2PN precession equations, and captures the precessional motion of the binary orbit around the direction of the total angular momentum~\cite{Apostolatos:1994} through in-plane spin components~\cite{Schmidt:2010it,Schmidt:2012rh,Schmidt:2014iyl}.

Beyond inspiral and PN considerations, spin dynamics also affect the merger regime which can dominate the observed signal from massive BBHs.
NR simulations show that the direction of GW emission keeps changing, i.e., precessing, during the merger~\citep{OShaughnessy:2012iol} and further affects the ringdown mode content~\cite{Siegel:2023lxl,Zhu:2023fnf,Hamilton:2021pkf,Hamilton:2023znn}.
Parameter inference based on simulated data further shows that spin constraints can be achieved from massive binaries with merger-dominated signals though they are typically weaker than mass constraints~\cite{Biscoveanu:2021nvg,Xu:2022zza}. 

Although the majority of GW signals observed are consistent with vanishing $\chieff$ and uninformative $\chip$~\cite{GWTC2.1,GWTC3}, notable exceptions exist.
These include some of the most massive BBHs detected, with a significant portion of the signal corresponding to the merger.
GW190521~\cite{GW190521_detection, GW190521_astro} is the most massive confidently detected BBH to date. 
While consistent with $\chieff=0$, its $\chip$ is constrained to large values, corresponding to large in-plane spins~\cite{GW190521_detection, GW190521_astro}.
Under the interpretation of a quasicircular BBH, the measurement is attributed to the suppression of the final pre-merger cycle due to precessional motion~\cite{Miller:2023ncs}. 
However, the short duration of the signal makes it subject to alternative explanations, like an eccentric~\cite{Romero-Shaw:2020thy,Gayathri:2020coq} or highly-unequal mass binary~\cite{Nitz:2020mga}, or a head-on collision~\cite{Gamba:2021gap}.
GW200129 displays similar spin properties~\cite{GWTC3,Hannam:2021pit}, but its interpretation is complicated by the fact that an instrumental glitch overlapped the portion of the signal that drives the $\chip$ measurement~\cite{Payne:2022spz,Macas:2023wiw}.
GW191109, notable for its negative $\chieff$~\cite{GWTC3}, also has an informative $\chip$ measurement, but also overlaps with a glitch~\cite{Udall:2024ovp}.
Further events have $\chip$ values that are informatively \textit{small}~\cite{GWTC3}, while a total of 10 events exclude $\chieff=0$ at 99\% credibility~\cite{GW151226,GW190412,GWTC2.1,GWTC2,GWTC3}
At the population level, the distributions of $\chieff$ and $\chip$ further point to the existence of BBHs that are precessing~\cite{LIGOScientific:2020kqk,KAGRA:2021duu}. 

The above discussion and most BBH spin constraints in the literature are based on $\chieff$ and $\chip$ (and its variants~\cite{Gerosa:2020aiw,Thomas:2020uqj}). 
However, both parameters are motivated by PN dynamics and their interpretation is only fully valid in the inspiral regime.
Though this does not invalidate their use to express constraints from merger-dominated events, it does suggest that they might not fully capture the imprint of spins on merger signals.
For example, in the case of non-precessing BBHs, perturbation theory suggests that post-merger emission is characterized by the remnant's mass and spin, for which there exist approximate fitting formulae that depend on the pre-merger masses and spin amplitudes~\cite{Healy:2018swt,Cheung:2023vki}.
Alternative merger-motivated spin parametrizations would aid in theoretically understanding the merger dynamics and devise more efficient parameter estimation methods, for example through re-parametrized sampling, e.g., Ref.~\cite{Roulet:2022kot}. 
Moreover, since spin constraints drive influential astrophysical conclusions, morphologically understanding their imprint on BBH mergers could safeguard against systematics from waveform models~\cite{Varma:2021csh}, instrumental glitches~\cite{Payne:2022spz,Udall:2024ovp}, or alternative interpretations~\cite{Romero-Shaw:2022fbf}. 
   
In the absence of simple analytic expressions for the merger dynamics, we propose an alternative approach {to study the correlation structure of BBH parameter space}. 
Efficient parametrizations typically simplify a problem by tracking symmetries and parameter correlations.
For example, the fact that the chirp mass is the best measured mass combination is equivalent to the fact that a small change in the chirp mass results in a very different signal. 
Two signals with different chirp masses therefore have a low match, i.e.,~normalized inner product.
If we had no knowledge of the chirp mass, its existence and formula could be identified by studying how signals differ as their parameters change or by examining the structure of the posterior for the component masses.
Such an approach is based on the ability to generate signals for various parameters, a situation that exactly corresponds to merging BBHs: using waveform models, we can simulate signals for any parameter combination even though we lack simple equations for the binary dynamics. 

In this work, we apply the above idea to spin parameters. 
Restricting to the seven-dimensional space of mass ratio and spins, we compute waveform mismatches between signals generated by BBHs with different parameters.
Mismatches are evaluated by extensions of the neural network constructed in~\citet{Ferguson:2022qkz} that maps binary parameters directly to signal mismatches, bypassing the computationally-inefficient steps of waveform generation and inner-product calculation.
Compared to that study, we enhance the network training data to incorporate higher-order modes, and additionally train networks at different values for the binary total mass so as to target different parts of the signal (inspiral versus merger).
In this first study, we restrict to systems with fixed values of the total mass. We consider the spin degrees of freedom that are aligned with the orbital angular momentum and those in the orbital plane separately, and use a Gaussian fit to map directions in the parameter space along which mismatches remain low.
We interpret these directions as ``spin correlation" directions, and study how they change for systems with different binary total mass, i.e.,~signals that are more or less merger-dominated.

We verify that this approach can recover the known imprint of $\chieff$ and mass ratio on low-mass, inspiral-dominated systems.
Increasing the aligned spin or decreasing the mass ratio leads to longer signals~\cite{Campanelli:2006,Healy:2018swt}, therefore the two parameters are positively correlated; there exists a prescription for increasing both such that the signal remains approximately unaltered.
This correlation is distinct from the $\chieff$-mass ratio correlation observed in parameter estimation~\cite{Cutler:1994ys,Poisson:1995ef,Baird:2012cu,Ng:2018neg,Fairhurst:2023idl}, since we fix the system total mass. 
As the total mass of the system increases and less of the inspiral is observable, both the shape and the strength of the $\chieff$-mass ratio correlation are altered.
The slope of the correlation in $\chieff$-mass ratio space decreases.
A smaller change in spin is required to cancel out a change in the mass ratio and thus the former's imprint on the signal is stronger.
Simultaneously, the strength of the correlation, quantified by the Gaussian fit covariance matrix, is weakened.
An analytic fit to the $\chieff$-mass ratio correlation as a function of the total mass provides a phenomenological quantitative description of the spin imprint that extends to merger-dominated signals.
Correlations between the in-plane spins and mass ratio also become less steep and strong as the total mass increases, effectively disappearing at a total mass of $270\,M_{\odot}$.

{Traditionally, BBH parameter-space exploration is done \textit{locally} around a given parameter point, either in full generality using parameter estimation (PE) with sampling~\cite{Ashton:2018jfp} or approximately through the use of the Fisher information matrix (FIM)~\cite{Owen:1995tm}.
Our method is technically related to the FIM, which is approximately the \textit{inverse} covariance matrix, whereas we study the eigen-directions of the covariance matrix.
More importantly, we extend the local analysis by tracing across a given parameter space to reveal \textit{non-local} correlation paths.}\footnote{{By ``non-local" we mean not restricted to be close to the original point of interest. For example, some paths span the full allowed region of possible spins and/or mass ratio.}} {Unlike traditional sampling, which probes correlations through computationally intensive posterior sampling, our approach efficiently explores a structured subset of parameter space informed by the local correlations. Each point in one of our correlation maps approximates what would otherwise require a full PE analysis, making the otherwise computationally prohibitive exploration of the BBH intrinsic parameter space feasible through mismatches~\cite{Hannam:2013uu} and a neural network~\cite{Ferguson:2022qkz}.}

In full generality, all 15 parameters of a quasi-circular BBH affect how the signal appears in a detector and thus if spins can be measured. 
To keep the problem manageable, we break down the full, complicated imprint of spins into its influence on reduced-dimension parameter-spaces. 
We restrict to intrinsic parameters only, fixing all extrinsic degrees of freedom (e.g., {inclination}, {sky position}, and distance). We further break down the space by fixing total mass and analyzing aligned-spin and precessing-spin systems separately.
For precessing systems, we restrict to single-spin configurations. 
This structured decomposition breaks the problem into manageable pieces and provides a first step toward characterizing mass–spin correlations in the merger regime, where such correlations remain poorly understood quantitatively. 
Our methodology thusly sets the groundwork for future studies in fuller parameter spaces lacking analytical descriptions.

The rest of the paper is organized as follows. 
In Sec.~\ref{sec:methods}, we present the details of the methodology including notation, the mismatch-predicting networks, and the correlation mapping algorithm.
In Sec.~\ref{sec:inspiral}, we apply the mapping algorithm to low-mass systems and aligned-spin degrees of freedom, while the aligned-spins of more massive systems (merger correlations) are considered in Sec.~\ref{sec:merger}.
In Sec.~\ref{sec:precession} we consider in-plane degrees of freedom for all masses.
Finally, our conclusions are given in Sec.~\ref{sec:conclusions}.

\section{Methodology}\label{sec:methods}

In this section, we describe the approach to exploring correlations in the parameter space of BBH signals. 
In Sec.~\ref{sec:binaries}, we introduce parametrizations used for a binary's masses and spins and establish notation.
In Sec.~\ref{sec:network}, we discuss the neural networks that map binary parameters to mismatches.
We detail the mapping algorithm that tracks low-mismatch directions along the parameter space in Sec.~\ref{sec:mapping correlations}.

\subsection{Parametrizing quasicircular binaries}
\label{sec:binaries}

Quasi-circular BBHs are characterized by $8$ intrinsic parameters: $2$ masses and $6$ components of the spin vectors.\footnote{An additional 7 extrinsic parameters describe a binary's location and orientation. 
{As these affect the waveform through known analytic expressions (e.g., antenna patterns)~\cite{Maggiore:2007ulw}, we work in the geocenter frame and focus on intrinsic parameters.} }
Masses can be expressed through the total mass $M=m_1+m_2$, the mass ratio $q=m_1/m_2\geq1$, and the symmetric mass ratio $\eta=q/(q+1)^2\in (0,0.25]$, where $m_i$, $i\in\{1,2\}$ are the component masses.
A symmetric mass ratio of $\eta=0.25$ corresponds to an equal-mass BBH; increasingly large $q$ map to increasingly small $\eta$.

The dimensionless component spin vectors $\vec{\chi}_{1}, \vec{\chi}_{2}$ are typically expressed in a frame whose $z$-axis denotes the direction of the Newtonian orbital angular momentum ${\hat{L}}$. 
The frame is defined at specific reference time or frequency; here we use a reference of 20\,Hz{, and fix the inclination to be face-on ($\iota = 0$), which maximizes the amplitude of the dominant $(2,2)$ mode, corresponding to the loudest and most likely observed signals. We note that care must be taken when interpreting results for precessing systems: since precession enters through inclination-dependent mode mixing~\cite{CalderonBustillo:2016rlt}, correlations may vary at other inclinations~\cite{Miller:2025eak}.}
Each vector has a magnitude $\chi_i$, polar (tilt) angle $\theta_i$ relative to ${\hat{L}}$, and azimuthal angle $\phi_i$ with respect to the $x$-axis, defined as the line of separation between the two BH.
Equivalently, $\vec\chi_i$ can be expressed in Cartesian components $\left(\chi_{ix},\chi_{iy},\chi_{iz}\right)$.

Motivated by PN equations, the $6$ spin degrees of freedom can be re-packaged into two effective parameters.
The effective aligned spin $\chieff$ is the mass-weighted average of the component spins along ${\hat{L}}$~\cite{Racine:2008qv, Ajith:2009bn}:
    \begin{equation}\label{eq:chieff}
        \chieff = \frac{q \chi_1 \cos\theta_1 + \chi_2 \cos\theta_2}{q+1} \in (-1,1)\,.
    \end{equation}
This quantity reduces to the leading-order spin-orbit term in the GW phase in the equal-mass limit and it is conserved under precession to at least the 2PN order~\cite{Racine:2008qv}.
When the BH spins are misaligned with ${\hat{L}}$, the orbital plane precesses about the direction of the total angular momentum~\cite{Apostolatos:1994}.
The effective precessing spin $\chip$ is the mass-weighted in-plane spin component motivated by the precession equations~\cite{Schmidt:2010it,Schmidt:2012rh,Schmidt:2014iyl} 
    \begin{equation}\label{eq:chip}
    \begin{split}
        \chip & = \max \left(\chi_1 \sin \theta_1, \left(\frac{4 +3q}{4q^2+3 q}\right)  \chi_2 \sin \theta_2 
    \right) \\ & \in [0,1)\,.
    \end{split}
    \end{equation}

Since we are interested in the imprint of binary parameters on the detected signal, all parameters are quoted in the detector frame.
    
\subsection{Waveform mismatch with neutral networks}
\label{sec:network}

We quantify the similarity between the signals emitted by two BBHs $h_1,h_2$ with the \textit{mismatch}~\cite{Owen:1995tm,Mohanty:1997eu} 
    \begin{align}
        {\cal{MM}} 
        & \equiv 1-\max _{t, \phi} \frac{\left\langle h_1 \mid h_2\right\rangle}{\sqrt{\left\langle h_1 \mid h_1\right\rangle\left\langle h_2 \mid h_2\right\rangle}}\in[0,1]\,,
        \label{eq:mismatch}
    \end{align}
where maximization is over relative time $t$ and phase $\phi$ shifts and the inner product is
    \begin{equation}\label{eq:inner_product}
        \left\langle h_1 \mid h_2\right\rangle=2 \int_{f_0}^{\infty} \frac{\tilde{h}_1^* \tilde{h}_2+\tilde{h}_1 \tilde{h}_2^*}{S_n} \,d f\,,
    \end{equation}
where tildes denote Fourier transforms, $S_{n}$ is the one-sided noise power spectral density, and $f_0=20\,$Hz is the starting frequency that corresponds to the detector low-frequency cutoff.
In what follows, we take $S_{n}\sim \rm constant$, i.e.,~a flat noise spectrum.
The mismatch is normalized such that ${\cal{MM}}=0$ represents identical waveforms and ${\cal{MM}}=1$ indicates orthogonal waveforms. 

Directly calculating enough mismatches to map parameter correlations in a high-dimensional parameter space is computationally expensive. 
Instead we turn to~\citet{Ferguson:2022qkz} in which a neural network is trained on the Simulating Extreme Spacetimes (SXS) NR waveform catalog~\cite{Boyle:2019kee}.
Originally constructed to identify regions of the parameter space that are sparsely covered by the catalog, the network provides a map between two sets of intrinsic BBH parameters $\boldsymbol{\lambda}\equiv\{\eta, \vec{\chi}_1, \vec{\chi}_2\}$ and the mismatch between their waveforms.
Mismatches refer to geocenter waveforms $h_i$ that have not been projected onto a detector network and thus focus on the intrinsic binary dynamics.
All signals are evaluated at a constant total mass determined by the length of the simulations.

\begin{figure}
    \centering
    \includegraphics[width=\linewidth]{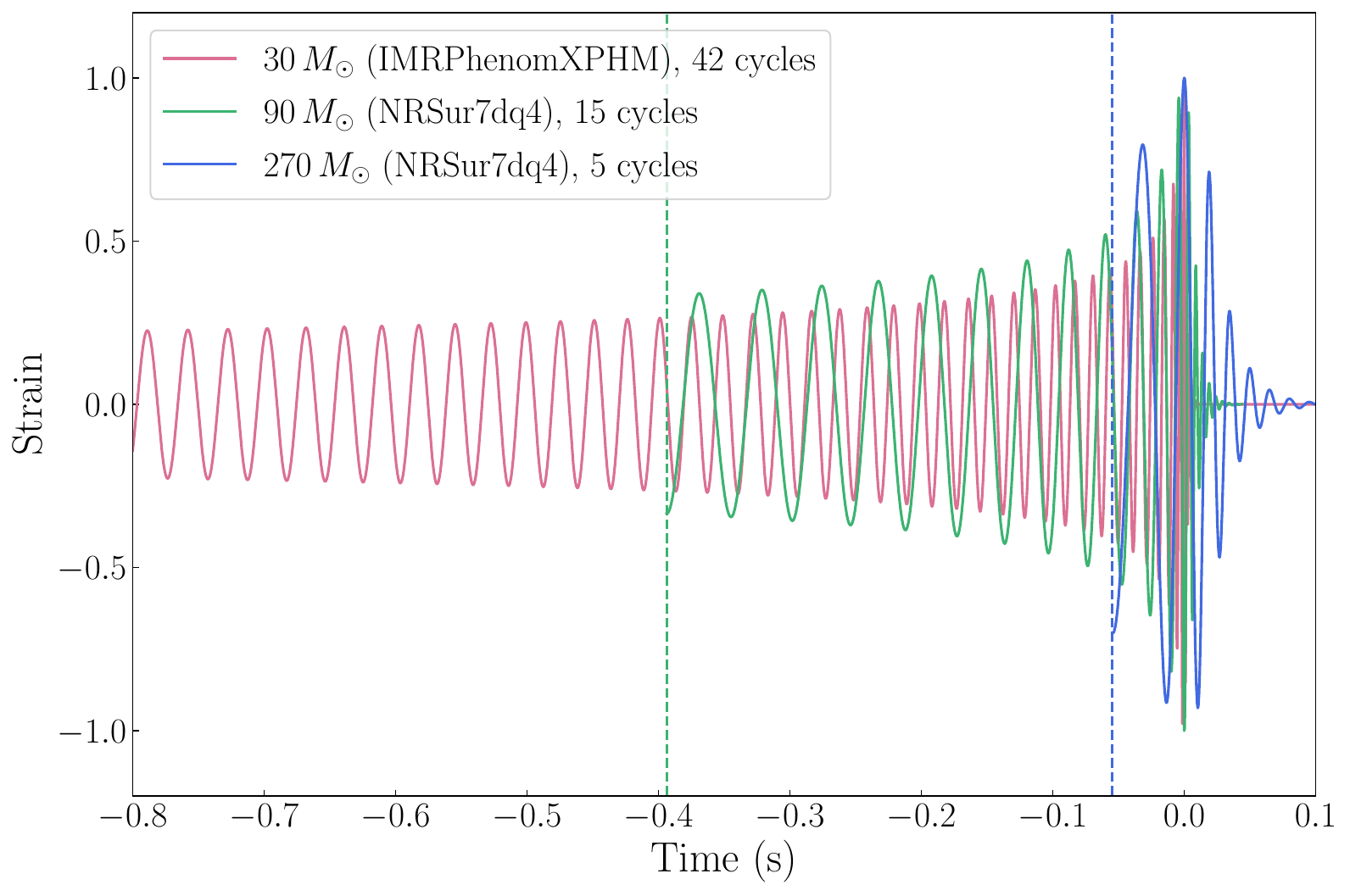}
    \caption{Scaled waveforms for equal-mass BBHs with zero spins at different total masses: $30 \, M_\odot$ (red), $90 \, M_\odot$ (green), and $270 \, M_\odot$ (blue) evaluated from 20\,Hz. Dashed lines indicate the start of the waveform and the number of cycles is given in the legend.
    }
    \label{fig:waveforms}
\end{figure}

We extend this network~\cite{Ferguson:2022qkz} as detailed in App.~\ref{app:network}.
Firstly, we include higher-order radiation modes and precession effects.
Secondly, as we are interested in exploring the impact of spins on different portions of the signal, we consider three values of the total mass: $30\,\Msun$, $90\,\Msun$, and $270\,\Msun$, training a separate network for each value.
The former corresponds to low-mass BBHs whose observed signal is dominated by the inspiral phase, the second to signals where both the late inspiral and the merger are observable, and the third to GW190521-like signals where only a handful of merger cycles are observed. 
For reference, Fig.~\ref{fig:waveforms} shows equal-mass nonspinning waveforms for each total mass.
Thirdly, rather than the SXS catalog, we train the network with waveform models. \textsc{NRSur7dq4} is a surrogate to NR with accuracy comparable to NR simulations~\cite{Varma:2019csw}. 
We therefore use it for the $90\,\Msun$, and $270\,\Msun$ networks.
Due to its finite length, however, \textsc{NRSur7dq4} cannot model the full signal from $30\,\Msun$ BBHs, for which we instead use the phenomenological \textsc{IMRPhenomXPHM}~\citep{Pratten:2020ceb} model.
Based on \textsc{NRSur7dq4}'s regime of validity, all networks are trained in the region $q\leq 6$ (equivalently $\eta > 6/49 \simeq 0.12$), $\chi_i<1$, $\theta_i\in[0,\pi]$, and $\phi_i\in[0,2\pi]$.

\subsection{Mapping parameter correlations}
\label{sec:mapping correlations}

\begin{figure*}[]
    \includegraphics[width=\textwidth]{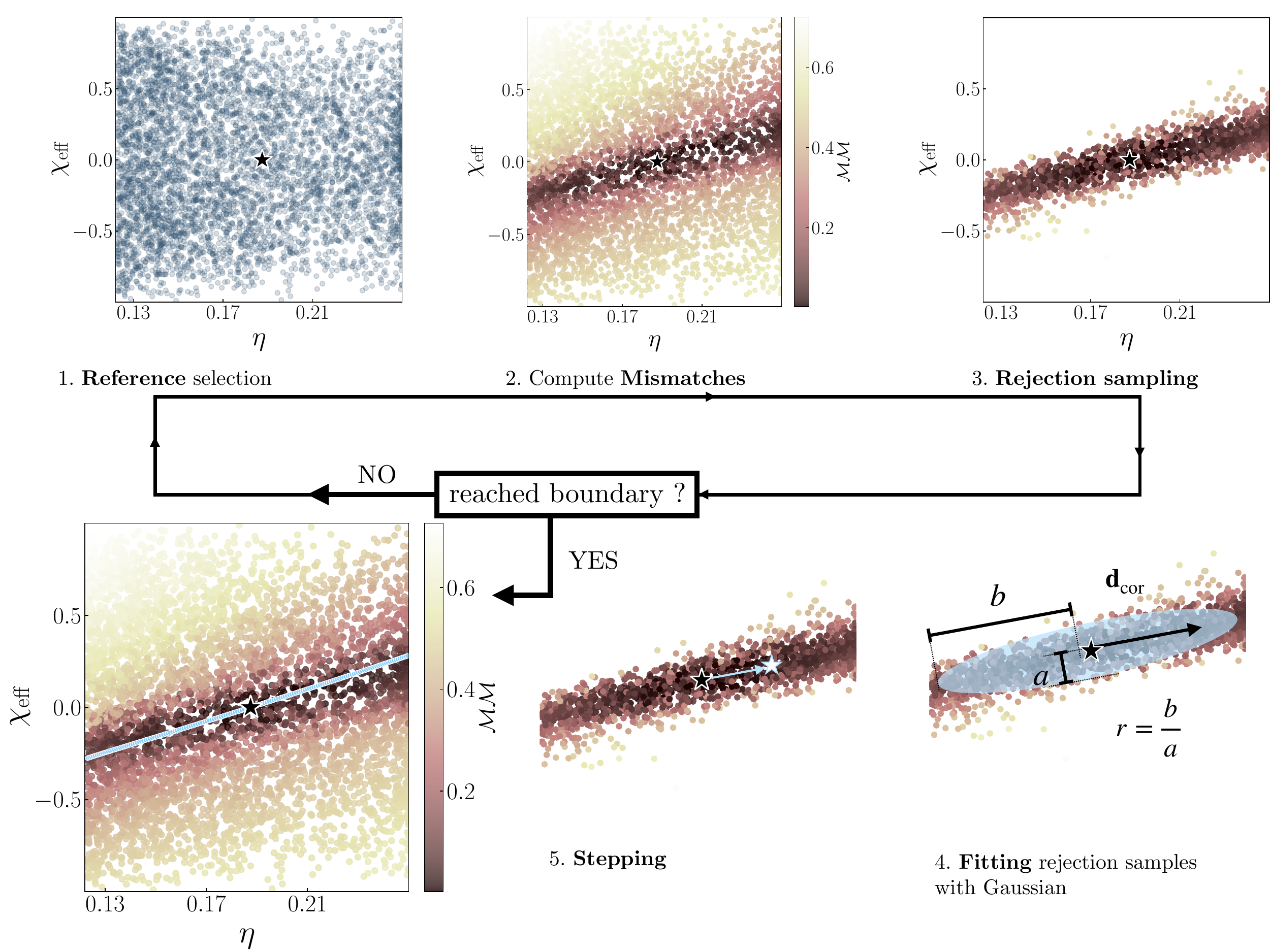}
    \caption{Workflow diagram for the correlation mapping algorithm described in Sec.~\ref{sec:mapping correlations}. This demonstration corresponds to mapping correlations in the 2D effective space of $\boldsymbol{x^{\rm(eff)}}=[\eta,\chieff]$ and the 90\,$M_{\odot}$ network. 
    1.~Reference point (black star) in a randomly sampled parameter space (blue dots); 2.~Mismatch with respect to the reference point (color map); 3.~Rejection sampling based on Eq.~\eqref{eq:weights}; 4.~The fitted 1-$\sigma$ contour, the maximum variance eigendirection of the covariance matrix (vector) $\boldsymbol{d}_{\mathrm{cor}}$, and the correlation strength $r$; 5.~Update the reference point along $\boldsymbol{d}_{\mathrm{cor}}$ per Eq.~\eqref{eq:stepping}. 
    We iterate through steps 1--5 until we reach a parameter space boundary in one of the dimensions over which we are mapping (in this visualization, the upper limit of $\eta$). 
    The total path taken in the $\chieff-\eta$ space is shown in the final panel (bottom left) together with the mismatches against the reference point.
    }
    
    \label{fig:flowchart}
\end{figure*}

We define parameter correlations as regions in the parameter space that result in similar waveforms. 
To identify these regions, we define the \emph{parameter space} that consists of $N$ draws of BBH parameters uniformly in symmetric mass ratio, spin magnitude, and spin directions within their training region: 
\begin{equation}
    {\cal{PS}}=\left\{\boldsymbol{\lambda}_i \mid \boldsymbol{\lambda}_i=\left\{\eta_i,\vec{\chi}_{1i},\vec{\chi}_{2i}\right\},\, \forall \, i \in N\right\}\,.
\end{equation}
Parameter correlations can be traced with different parameter subsets and we consider the following.
\begin{enumerate}
    \item \textbf{Effective} parameter space: Knowledge of the form of $\chieff$ and that $\eta$ (rather than $q$) appears in the GW phase, suggests directly mapping correlations in the reduced $2$-dimensional $\boldsymbol{x^{\rm (eff)}}=[\eta,\chieff]$ space.
    In this space, the full parameter set $\boldsymbol{\lambda}$ is not fully determined; we select $\boldsymbol{\lambda}=\left\{\eta, \vec{\chi}_1=(0,0,\chieff),\vec{\chi}_2=(0,0,\chieff)\right\}$. 
    \item \textbf{Aligned} parameter space: When we instead consider generic aligned-spin correlations (e.g.,~a generalization of $\chieff$), we map the $3$-dimensional space of $\boldsymbol{x^{\rm (al)}}=[q, \chi_{1 z}, \chi_{2 z}]$. 
    The full parameter set $\boldsymbol{\lambda}$ is now fully determined by $\boldsymbol{x^{\rm (al)}}$.
    \item \textbf{Precessing} parameter space: Correlations between the in-plane spin degrees of freedom correspond to the $5-$dimensional space of $\left[\eta, \chi_{1x}, \chi_{2x}, \chi_{1y}, \chi_{2y}\right]$. 
    However, since two-spin precession effects~\cite{Purrer:2015nkh,Chatziioannou:2018wqx} and the imprint of the in-plane azimuthal angle~\cite{Kalaghatgi:2020gsq} are weak,\footnote{We have verified both conclusions by finding no correlations in preliminary results involving $\vec{\chi}_2$ and $\phi_1$.} we restrict to a single precessing spin along the binary $x$-axis and the 2-dimensional space of $\boldsymbol{x^{\rm (pre)}}=\left[\eta, \chi_{1}\right]$, thus setting the secondary spin and other components of the primary spin to zero. 
    The full parameter set $\boldsymbol{\lambda}$ is then fully determined by $\boldsymbol{x^{\rm (pre)}}$.
\end{enumerate}
In summary, each $\boldsymbol{x}$ (effective, aligned, precessing) is a subset of $\boldsymbol{\lambda}$ with some parameters fixed to a constant value.

Given a choice of parameters $\boldsymbol{x}$ within which we look for correlations and the appropriate neural network \texttt{NN}, we map the correlation directions iteratively.
The process is shown schematically in Fig.~\ref{fig:flowchart} for the $\chieff-\eta$ correlation and the $90\,M_{\odot}$ network.
\begin{enumerate}
    \item {\bf Reference}: We select a reference point in the appropriate parameter space (effective, aligned, precessing) $\boldsymbol{\lambda}_{j} $ and corresponding $\boldsymbol{x}_{j} $ around which we look for correlations. 
    \item {\bf Mismatches}: Using the neural network, we compute the mismatch $\cal{MM}$ between the reference and all other points in ${\cal{PS}}$:
    \begin{equation}
    \begin{split}
    \qquad{\cal{M}} & {\cal{M}} =\\ \qquad &\left\{{\cal{MM}}_i \mid {\cal{MM}}_i= \texttt{NN}\left(\boldsymbol{\lambda}_i, \boldsymbol{\lambda}_{j}\right), \forall \boldsymbol{\lambda}_i \in {\cal{PS}}\right\}\,.
    \end{split}
    \end{equation}
    A colormap of the mismatches reveals the correlation direction in Fig.~\ref{fig:flowchart}; here the region of lowest mismatch (darkest color) is a diagonal path from negative $\chieff$-low $\eta$ to positive $\chieff$-high $\eta$.
    
    \item {\bf Rejection Sampling}:
    Each point in $\boldsymbol{\lambda}_i$ receives a weight $w_i$ inspired by the dependence of the likelihood on the mismatch~\cite{vanderSluys:2007st,vanderSluys:2008qx,Veitch:2008,Veitch:2014wba,Ashton:2018jfp}
    \begin{equation}
    w_i=\exp \left(-\frac{\rho^2  {\cal{MM}}_i^2}{2}\right)\,,
    \label{eq:weights}
    \end{equation}
    where $\rho$ corresponds to the signal-to-noise ratio (SNR) in a parameter estimation setting. Higher SNR signals result in posteriors with mismatches more tightly clustered around $0$. 
    Given these weights, we perform rejection sampling to retain points with low mismatch. 

    \item {\bf Fitting}: 
    We define the correlation direction $\boldsymbol{d}_{\text {cor}}$ (which has the same dimensionality as $\boldsymbol{x}$) as the direction of maximum variance--equivalently, maximum eigenvalue, i.e.,~mismatch values remain close to $0$. The direction $\boldsymbol{d}_{\text {cor}}$ and its eigenvalue $\lambda_\mathrm{cor}$ are computed with a Gaussian fit; see App.~\ref{app:PCAGMM} for details, where we also confirm that we obtain similar results with a principal component analysis. 

    To compare the correlation strength between different paths, we use the ratio of the semi-major to the semi-minor axis of the 1-$\sigma$ contour, $r$.
    The light blue ellipse in Fig.~\ref{fig:flowchart} represents this contour.
    The correlation strength is the square root of the condition number of the Gaussian covariance matrix, defined as the ratio of its largest and its smallest eigenvalue. We caution against over-interpreting the exact numerical value of $r$: it should be viewed as an order-of-magnitude estimate of the correlation strength, i.e., $r\sim\mathcal{O}(100)$ represents a strong correlation, $r\sim\mathcal{O}(10)$ is moderate, while $r\sim\mathcal{O}(1)$ is weak. It also depends on the dimensionality of the parameter-space in which correlations are measured.

    \item {\bf Stepping}: We define the next reference point as
    \begin{equation}
   \boldsymbol{x}_{j+1} =  \boldsymbol{x}_{j}+ \frac{\alpha \lambda_\mathrm{cor}}{\sqrt{D}} \boldsymbol{d}_{\text {cor }}\,,
   \label{eq:stepping}
    \end{equation}
    where $\alpha$ is a dimensionless free parameter that controls the size of the steps, and $D$ is the dimensionality of $\boldsymbol{x}$.
    \item {\bf Repeat} from \#1 and in the opposite direction until the parameter space boundary is reached. The list of points $\boldsymbol{x}_{j}$ map the correlation direction for the initial reference point $\boldsymbol{x}_{0}$.
    
\end{enumerate}

We tune free parameters to ensure that correlation directions are reliably recovered. 
Specifically, the total number of points $N$ in ${\cal{PS}}$ and $\rho$ are determined empirically such that we have a sufficient number of samples remaining after rejection sampling. 
We typically use $N = 100,000$, while $\rho$ is increased with the total mass and the number of dimensions $D$ to ensure at least 1\% efficiency in rejection sampling.\footnote{{The SNR cut focuses sampling on the low-mismatch region and provides a more robust and consistent selection than a hard mismatch threshold, which would otherwise require arbitrary initial configuration and mass-dependent tuning.}} 
The choice of $\alpha$ depends on the bounds of the parameter space we are mapping. 
We typically take $\alpha=1$ in the effective space and $\alpha=0.1$ for the aligned and precessing spaces.

The correlation mapping process is looking for \emph{paths}, i.e., 1-dimensional lines, across the parameter space. Network uncertainties may broaden the recovered correlations but do not impact their direction; Appendix ~\ref{app:network} shows percent level errors, while the correlations we are recovering below are at the $\mathcal{O}(10\%)$.
Since we restrict to a relatively small number of parameters (up to 3 for the aligned parameter space), this process still recovers non-trivial structure. Exploring correlations with more parameters (for example all 8 intrinsic degrees of freedom) can be achieved by generalizing the above method to higher-dimensional correlation hypersurfaces. 
Appendix~\ref{app:3Dsurface} further discusses this issue for the 3D aligned spin space. 
\section{Recovering Spin-Aligned Inspiral correlations at $30\,M_{\odot}$}
\label{sec:inspiral}

\begin{figure*}[]
    \includegraphics[width=\linewidth]{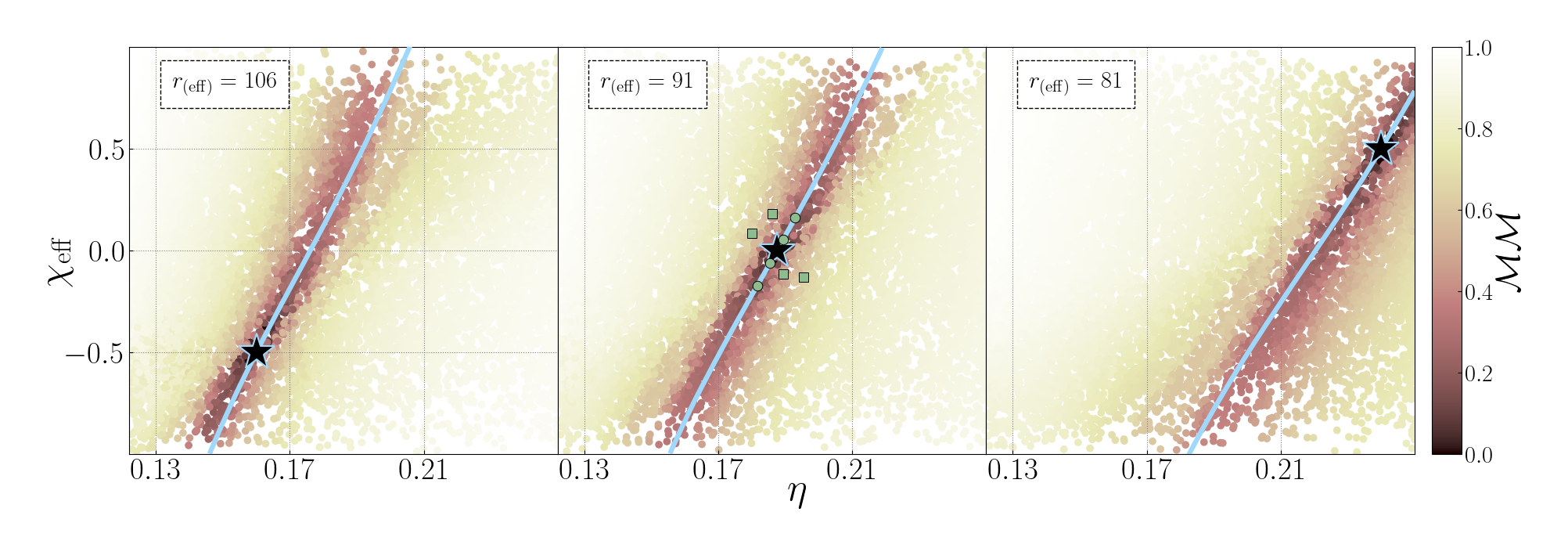} 
    \caption{Correlation path (blue line) in $[\eta, \chieff]$ for binaries with a total mass of $30\,M_{\odot}$ and different initial reference points: $(\eta, \chieff) = (0.16, -0.5), (0.19, 0), (0.24, 0.5)$ (left to right, star). 
    The colormap indicates the mismatch between the initial reference point and samples in the parameter space. 
    The correlation is visible by eye in the colormap and the computed paths track it consistently. 
    The correlation strength $r_{\rm{(eff)}}$, corresponding to the ratio of the semi-major to the semi-minor axis of the 1-$\sigma$ contour is given in-plot, confirming that the right-most system has the weakest correlation, as also evident from the colormap.
    Green circles and boxes in the middle panel denote the waveforms plotted in Fig.~\ref{fig:30_waveforms}.}
    \label{fig:2D_30_chieffeta}
\end{figure*}

We begin by exploring correlations in systems with a total mass of $30\,M_{\odot}$ where a major portion of the observed signal corresponds to the inspiral, c.f.~Fig.~\ref{fig:waveforms}.
Analytical knowledge from PN equations suggests that an appropriate spin combination for the aligned spin dynamics is $\chieff$, e.g., Refs.~\cite{Arun:2009,Ng:2018neg}. 
We therefore start by mapping correlations in the 2D space of $\boldsymbol{x^{\rm(eff)}}=[\eta,\chieff]$.
Figure~\ref{fig:2D_30_chieffeta} demonstrates how the correlation mapping algorithm of Sec.~\ref{sec:mapping correlations} recovers correlations in the $\chieff-\eta$ space.
Each panel corresponds to a different initial reference point $\boldsymbol{x^{\rm(eff)}}_0$.
We present $3$ systems with $(\eta, \chieff) = (0.16,-0.5), (0.19,0), (0.24, 0.5)$ (mass ratios of 4, 3, and 1.5 respectively), but obtain similar results for other configurations.
The colormap denotes the mismatch across the parameter space with respect to the initial reference point (star), where the white/yellow region indicates a region of high mismatch (dissimilar waveforms).
In all cases the recovered correlation path (light blue) agrees well with the region of low mismatches (maroon).
Moreover, regions of low mismatch are tightly concentrated around the correlation path, showing that the correlation itself is strong. 
For Fig.~\ref{fig:2D_30_chieffeta}, the correlation strength, i.e.,~the ratio of the semi-major to the semi-minor axis of the 1-$\sigma$ contour $r_{\rm{(eff)}}$ is 106, 91, 81 from left to right.
These numbers confirm the by-eye observation that the correlation is the least tight in the right panel, albeit not significantly.

\begin{figure}
    \includegraphics[width=\linewidth]{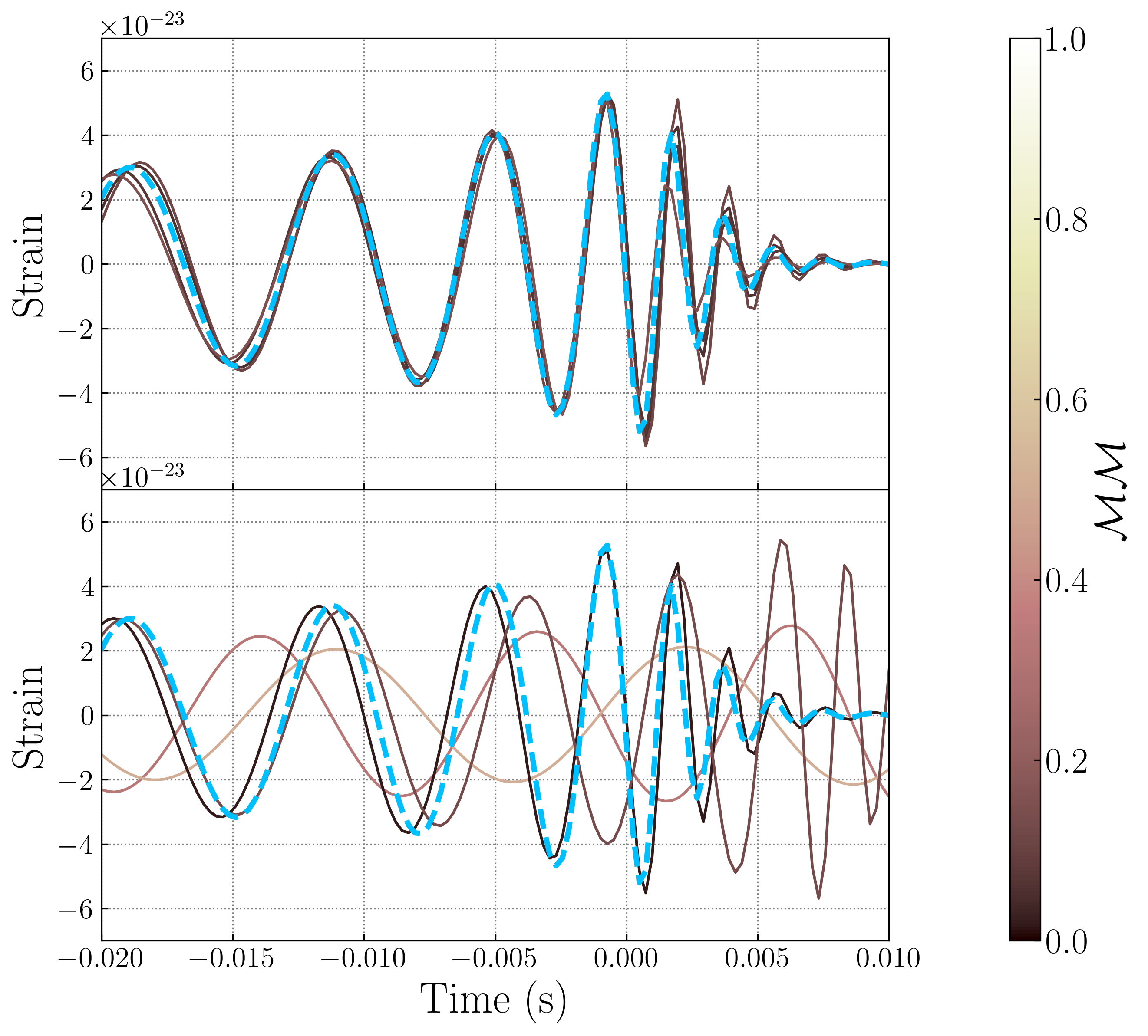}
    \caption{\textsc{IMRPhenomXPHM} waveforms demonstrating the $\chieff-\eta$ correlation for the reference point $(\eta,\chieff)=(0.19,0)$ from Fig.~\ref{fig:2D_30_chieffeta}.
    The blue dashed line is the reference point waveform. 
    The top panel shows waveforms along the correlation path of the reference point, colored by their mismatch with respect to the reference (green circles in Fig.~\ref{fig:2D_30_chieffeta}).
    These waveforms are visually similar and all have small mismatches with the reference waveform.
    The bottom panel shows random waveforms near the reference injection (green boxes in Fig.~\ref{fig:2D_30_chieffeta}). 
    We plot the waveforms with the minimum mismatch, i.e., minimized over time and phase shifts, with respect to the reference.}
    \label{fig:30_waveforms}
\end{figure}

A waveform demonstration of the correlation is provided in Fig.~\ref{fig:30_waveforms} for the reference point with $(\eta, \chieff)=(0.19, 0)$.
The top panel shows waveforms along the correlation path, while the bottom panel shows waveforms outside the correlation, all labeled by their mismatch and denoted in Fig.~\ref{fig:2D_30_chieffeta} with green dots and boxes respectively.
In the top panel, waveforms are by-eye similar to the reference waveform as expected from the low mismatch. 
Our analysis leverages this similarity to identify a correlation between $\chieff$ and $\eta$.

\begin{figure*}
    \includegraphics[width=\linewidth]{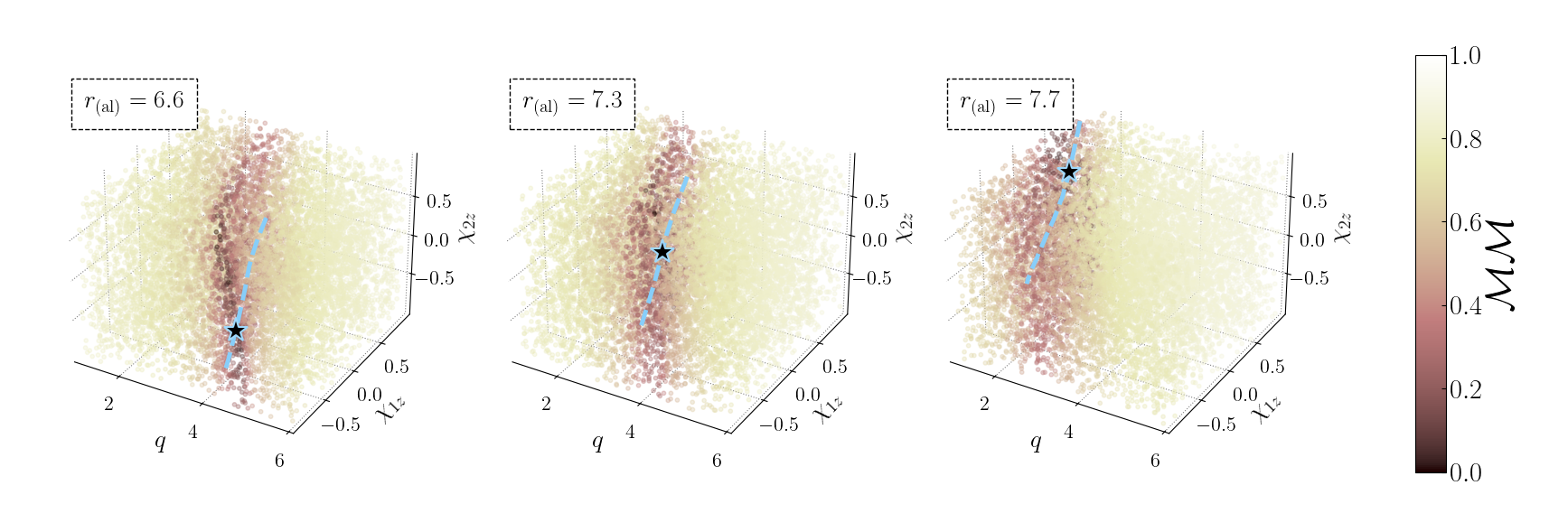}
    \caption{Correlation paths in $[q,\chi_{1z},\chi_{2z}]$ for binaries with a total mass of $30\,M_{\odot}$ and the same initial reference points (stars) as Fig.~\ref{fig:2D_30_chieffeta}: $(q, \chi_{1z},\chi_{2z}) = (4,-0.5,-0.5), (3,0,0), (1.5,0.5,0.5)$. Correlations are directly mapped in the 3D aligned spin space of $\boldsymbol{x^{\rm(al)}}=[q,\chi_{1z},\chi_{2z}]$.
    A coherent path is tracked across the parameter space, even though it has a higher dimensionality than strictly necessary. As a consequence the correlation strengths are lower than those of Fig.~\ref{fig:2D_30_chieffeta}.}
\label{fig:3D_30_qchi1zchi2z}
\end{figure*}

Having validated the correlation mapping algorithm in the effective space, we consider the more complex scenario in which we pretend that \textit{a priori} we do not know which aligned spin combination (i.e.,~$\chieff$) is most relevant. 
We therefore consider the full 3D aligned spin space with $\boldsymbol{x^{\rm(al)}}=[q,\chi_{1z},\chi_{2z}]$.
Figure~\ref{fig:3D_30_qchi1zchi2z} shows the correlation paths in $[q,\chi_{1z},\chi_{2z}]$, mapped directly in the 3D aligned spin space.
The smoothness of the paths indicates that a well-defined correlation does exist, though from this projection the lines do not appear to align with the darkest (smallest) mismatches, see App.~\ref{app:3Dsurface} for an explanation.
The recovered correlation has a reduced correlation strength of 6.6, 7.3, 7.7 from left to right.
The left panel of Fig.~\ref{fig:3D_30_90_270_projected} (discussed more later) shows the correlation paths projected into the 2D subspaces formed from $[q,\chi_{1z},\chi_{2z}]$.
\textit{If we had no knowledge of the analytical form of} $\chieff$\textit{, these correlation paths could aid us in motivating its formula}.
For example, as the correlation paths are smoother in $[q,\chi_{1z}]$ than the subspaces including $\chi_{2z}$, we would know that our effective spin parameter would more strongly depend on $\chi_{1z}$ than $\chi_{2z}$. 
This is true for $\chieff$: the primary spin is upweighted by $q$; see Eq.~\eqref{eq:chieff}.

\begin{figure}
    \includegraphics[width=\linewidth]{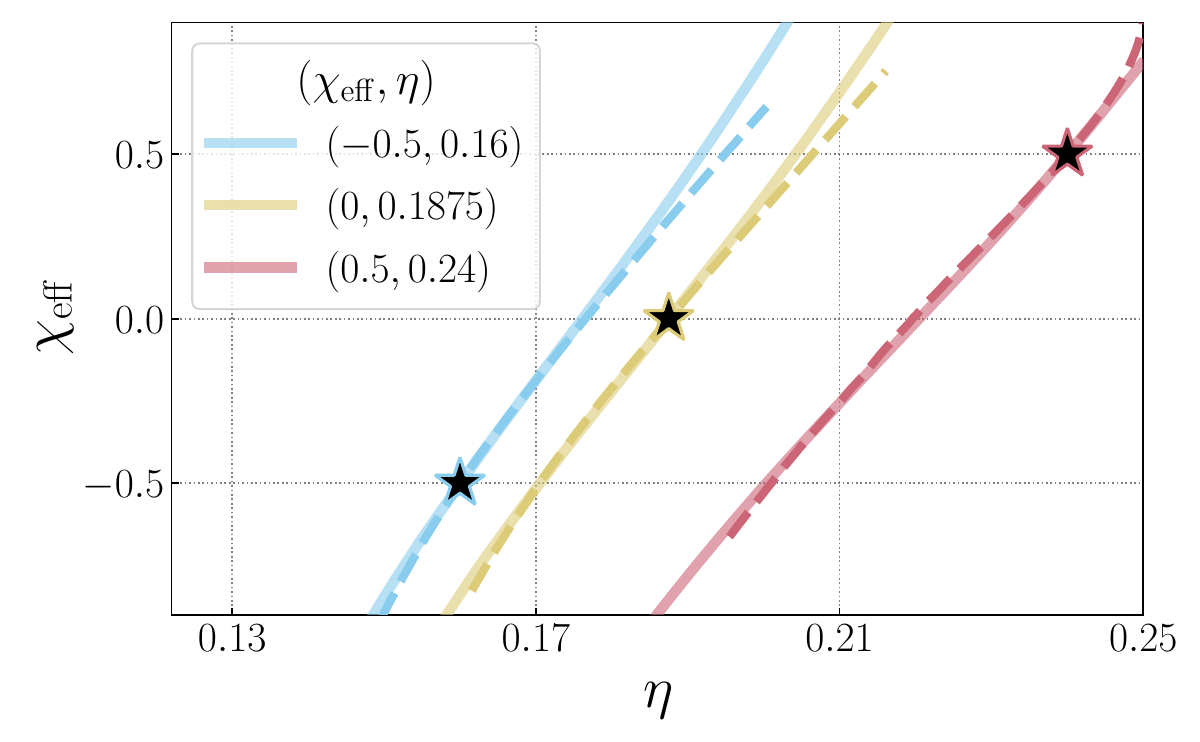}
    \caption{
    Correlation paths in $[\eta, \chieff]$ for binaries with a total mass of $30\,M_{\odot}$ and the same initial reference points (stars) as Fig.~\ref{fig:2D_30_chieffeta}. 
    We plot the 2D effective space of $\boldsymbol{x^{\rm(eff)}}=[\eta,\chieff]$ (solid; same as Fig.~\ref{fig:2D_30_chieffeta}) and the 3D aligned space of $\boldsymbol{x^{\rm(al)}}=[q,\chi_{1z},\chi_{2z}]$ (dashed; same as Fig.~\ref{fig:3D_30_qchi1zchi2z}) mapping that has then been projected to the $[\eta, \chieff]$ space.
    Correlation paths are consistent, thus validating the higher-dimensional correlation mapping.}
    \label{fig:2D3D_30_chieffeta}
\end{figure}

Figure~\ref{fig:2D3D_30_chieffeta} compares the correlation paths in $[\chieff,\eta]$ computed in different ways: 1.~directly from the 2D effective space $\boldsymbol{x^{\rm(eff)}}=[\eta,\chieff]$, and 2.~from the 3D aligned spin space $\boldsymbol{x^{\rm(al)}}=[q,\chi_{1z},\chi_{2z}]$ and then \textit{projected} to $[\chieff,\eta]$. 
For each reference system, the effective ($\boldsymbol{x^{\rm(eff)}}$; solid lines) and aligned ($\boldsymbol{x^{\rm(al)}}$; dashed lines) correlation mappings return consistent paths when projected to 2D.
Even when mapping a correlation in more dimensions than strictly necessary, our algorithm can recover equivalent structure.

The morphology of the correlation we recover, c.f., Fig.~\ref{fig:2D_30_chieffeta}, can be explained as follows.
For binaries with fixed total mass, both a larger $\chieff$ and more unequal masses increase the length of the waveform.
The former is because the binary has more angular momentum and energy to emit before merger, and the latter because energy emission is weaker.
Therefore the two are positively correlated: increasing both together leads to their effect approximately canceling out and the waveform remains approximately unchanged.
The result of Fig.~\ref{fig:2D_30_chieffeta} would have aided us in deducing this physical imprint of $\chieff$ and $\eta$ on the waveform, if we did not already know about it through PN equations.
Notably, this is not the same as the ``inspiral spin-mass ratio correlation"~\cite{Cutler:1994ys,Poisson:1995ef,Baird:2012cu,Ng:2018neg,Fairhurst:2023idl}, which instead leads to \emph{anti}-correlated $\chieff$ and $\eta$. 
That correlation is instead driven by the functional form of the 1.5PN phase term.
The compounding effect on the waveform length of changing $\chieff$ and $\eta$ in an anti-correlated fashion is counteracted by a change in the total mass.
Fixing the total mass instead completely alters the shape of the correlation.
Preliminary results with an expanded network that varies the total mass suggest that our methodology can recover the spin-mass ratio correlation of Ref.~\cite{Cutler:1994ys}; we leave further such explorations to future work.

\section{Aligned-spin correlations beyond the inspiral}\label{sec:merger}

As a binary approaches merger, the orbital velocity increases and the PN expansion becomes inaccurate.
Late inspiral and merger dynamics alter the signal evolution and change the spin correlations.
In this section, we explore spin correlations beyond inspiral-dominated signals.
We consider binaries with a total mass of $90\,M_{\odot}$ whose observable signal includes both the late inspiral and the merger, and $270\,M_{\odot}$ whose observable signal is overwhelmingly dominated by the merger.
Figure~\ref{fig:waveforms} shows example signals for reference.

\begin{figure*}
    \includegraphics[width=\linewidth]{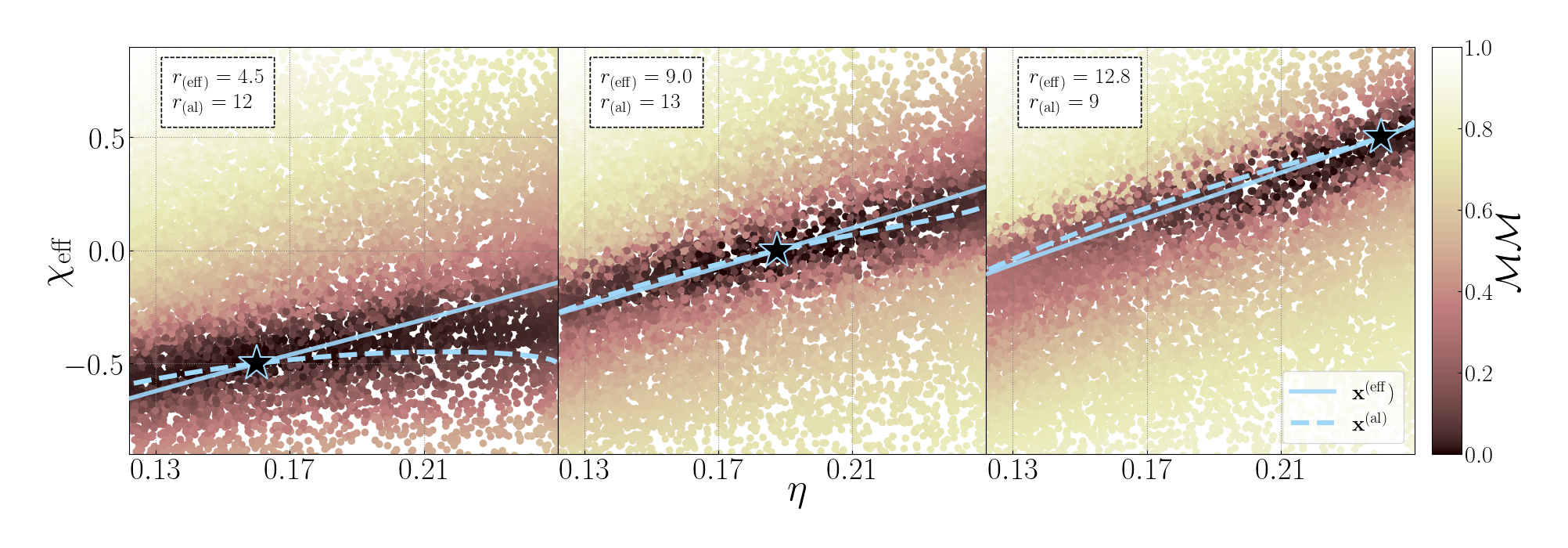}
    \caption{Similar to Fig.~\ref{fig:2D_30_chieffeta} but for the $90\,M_{\odot}$ network and signals where both the inspiral and the merger phase are observable. We also include in dashed the correlation paths from the 3D aligned space of $\boldsymbol{x^{\rm(al)}}=[q,\chi_{1z},\chi_{2z}]$ mapping that have then been projected to the $[\eta, \chieff]$ space. 
    The correlation paths are generally consistent, suggesting that $\chieff$ is still an appropriate aligned spin parameter at this total mass.
    The effective mapping correlation strength numbers are lower than those of Fig.~\ref{fig:2D_30_chieffeta}, reflecting the weakening of the $ \chieff{-}\eta$ correlation for heavier systems. The aligned spin mapping numbers are higher than the effective mapping ones, likely suggesting that a better spin-aligned parameter than $\chieff$ might exist. }
   \label{fig:2D3D_90_chieffeta}
\end{figure*}

With the PN expansion breaking down close to merger, it is no longer clear whether $\chieff$ is the appropriate spin aligned parameter.
However, as a first approximation it is reasonable to explore whether a correlation in $\chieff{-}\eta$ is still present, albeit in a weakened and/or altered form.
Figure~\ref{fig:2D3D_90_chieffeta} shows correlations recovered in the 2D effective space of $\boldsymbol{x^{\rm(eff)}}=[\eta,\chieff]$ with the $90\,M_{\odot}$ network. 
Firstly, some degree of correlation between $\chieff$ and $\eta$ exists: the colormap shows a clear structure and the mapping algorithm recovers coherent correlation paths.
Secondly, compared to the equivalent Fig.~\ref{fig:2D_30_chieffeta} for $30\,M_{\odot}$ signals, the correlation here is much \textit{weaker}, evident by the fact that low mismatches are spread out over a larger region around the correlation path (solid lines). 
Correspondingly, the correlation strength reduces to 4.5, 9.0, 12.8 from left to right.
Third, and again comparing to Fig.~\ref{fig:2D_30_chieffeta}, the correlation structure is \textit{altered}, i.e.,~has a different ``slope" in the $\chieff{-}\eta$ plane.

The altered and weakened $\chieff-\eta$ correlation for $90\,M_{\odot}$ signals motivates exploring correlations in the 3D aligned space of $\boldsymbol{x^{\rm(al)}}=[q,\chi_{1z},\chi_{2z}]$.
The correlation in the $\boldsymbol{x^{\rm(al)}}$ space projected to $[\eta,\chieff]$ is included in Fig.~\ref{fig:2D3D_90_chieffeta} (dashed lines).
The two paths, computed from the 3D aligned and the 2D effective spaces, are largely consistent, especially in the vicinity of the reference point. 
Differences emerge away from the reference point with the 3D paths being more curved, likely due to the larger flexibility of an extra dimension.
This discrepancy is the largest for the negative $\chieff$ unequal-mass test case (left panel).
Overall, $\chieff$ is still a relevant aligned spin parameter for systems with a total mass of $90\,M_{\odot}$.
The 2D effective mapping correlation strength numbers are lower than those of Fig.~\ref{fig:2D_30_chieffeta} for $30\,M_{\odot}$ systems, reflecting the weakening of the $ \chieff{-}\eta$ correlation that is also visible in the colormaps.
However, contrary to the $30\,M_{\odot}$ case, for 90\,$M_{\odot}$ binaries the 3D aligned spin correlation has a comparable (or in same cases larger) strength than the 2D effective one ($r_{\rm (al)} > r_{\rm (eff)}$). 
This likely indicates although $\chieff$ is an appropriate aligned spin parameter for signals without much visible inspiral, there might exist an even better one.

\begin{figure*}
    \includegraphics[width=\linewidth]{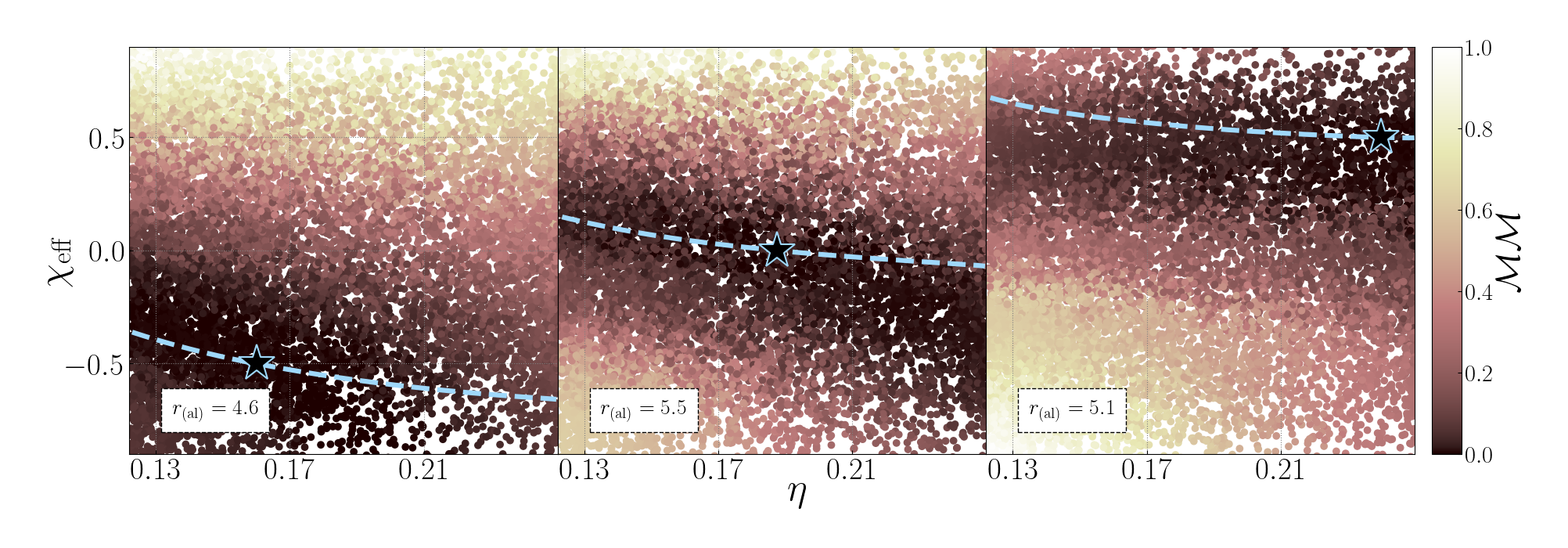}
    \caption{Similar to Fig.~\ref{fig:2D3D_90_chieffeta} but for the $270\,M_{\odot}$ network and signals where only the merger phase is observable.
    In contrast to Fig.~\ref{fig:2D3D_90_chieffeta}, we only include the correlation paths from the 3D aligned space of $\boldsymbol{x^{\rm(al)}}=[q,\chi_{1z},\chi_{2z}]$ mapping that have then been projected to the $[\eta, \chieff]$ space. 
    The correlation strength is lower than Fig.~\ref{fig:2D3D_90_chieffeta}, showing only weak trends in $[\eta, \chieff]$ for merger-dominated systems, and works in the opposite direction: $\chieff$ and $\eta$ are here \textit{anti-correlated}, albeit weakly.}
   \label{fig:2D3D_270_chieffeta}
\end{figure*}

The situation is different for the $270\,M_{\odot}$ signals.
Initial comparisons between the 2D effective and 3D aligned spaces revealed very different correlation paths.
As such, we only present 3D results, which are more generic.
Figure~\ref{fig:2D3D_270_chieffeta} shows results in the $[\eta, \chieff]$ space with correlation paths computed in the 3D aligned space and then projected. 
By eye the correlation seems to be almost completely gone with little structure apparent in the colormap.
The correlation path still roughly follows the region of low mismatch with a marginally negative slope.
The large extent of the low-mismatch region is reflected on the correlation strength: we report $r_{\rm{(al)}}=$ 4.6, 5.5, 5.1 compared to 12, 13, 9 for systems with identical spins and mass ratio, but a total mass of $90\,M_{\odot}$.

\begin{figure*}[]
\includegraphics[width=0.32\linewidth]{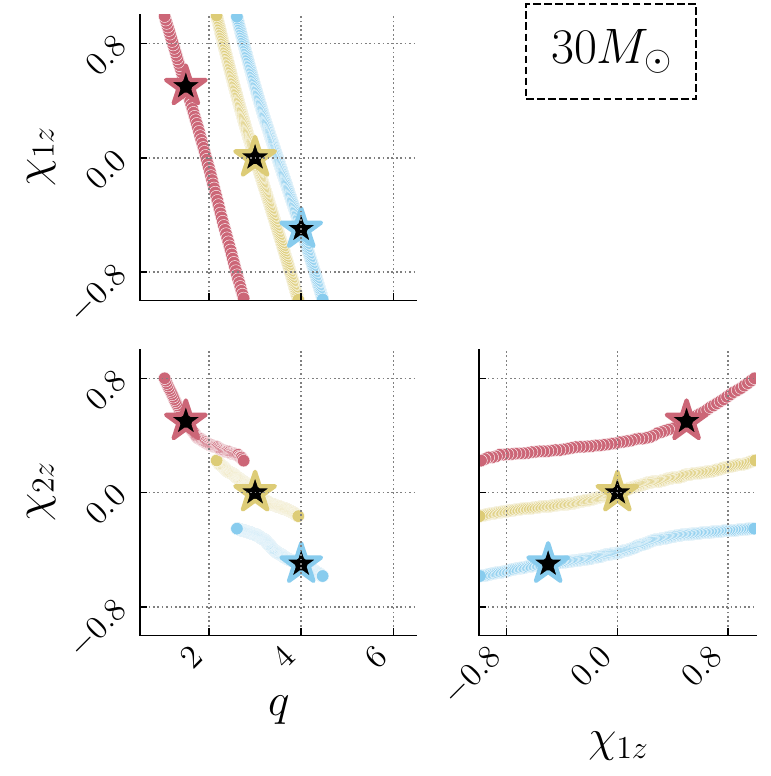} 
\includegraphics[width=0.32\linewidth]{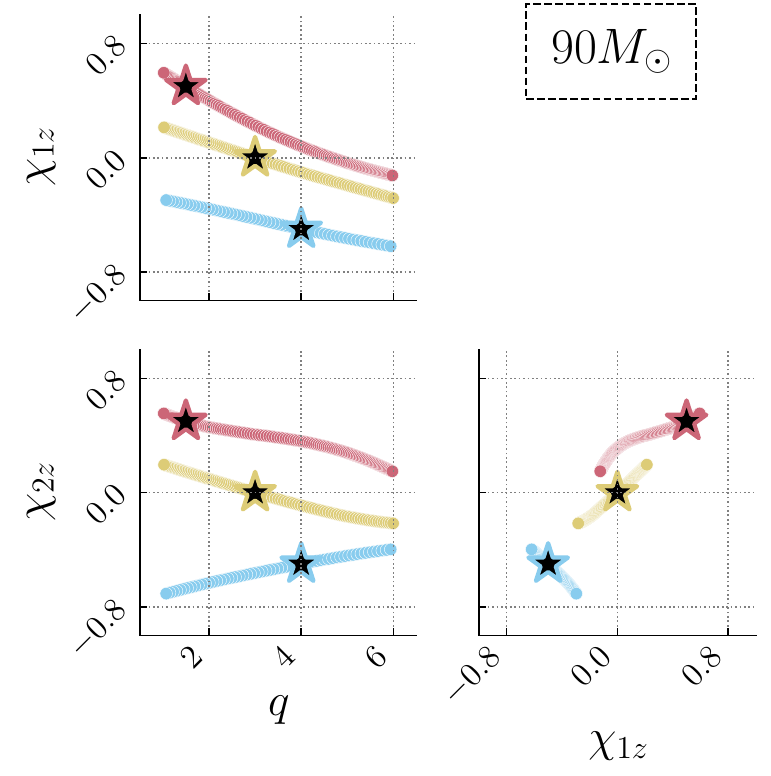} 
\includegraphics[width=0.32\linewidth]{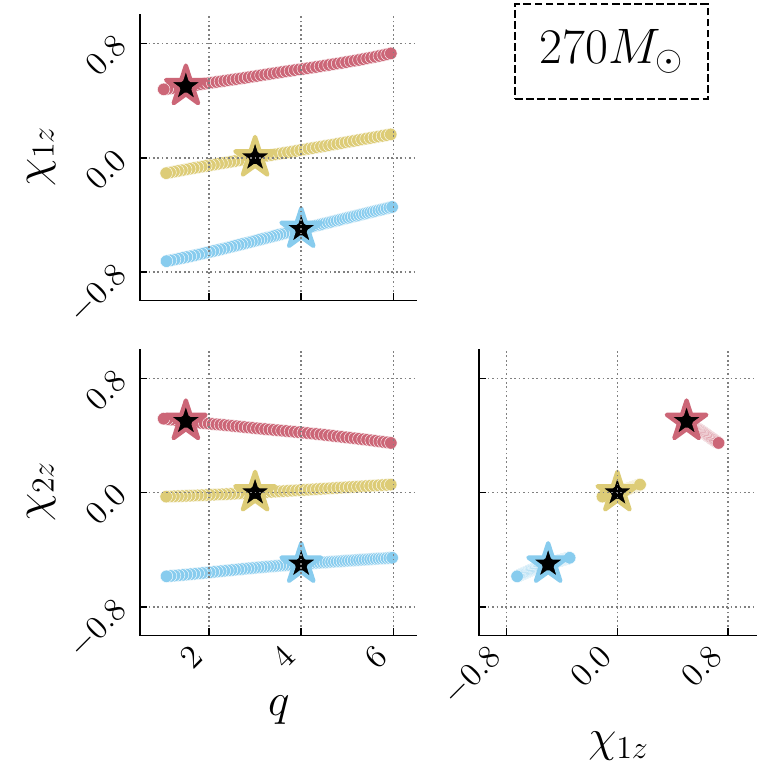} 
\caption{Correlation paths that have been constructed in the 3D aligned spin space of $[q,\chi_{1z},\chi_{2z}]$ and then projected in the 2D subspaces for the 30\,$M_{\odot}$ (left), 90\,$M_{\odot}$ (middle), and 270\,$M_{\odot}$ (right) networks. We show the same initial reference points as Fig.~\ref{fig:3D_30_qchi1zchi2z}, indicated by different colors. }
\label{fig:3D_30_90_270_projected}
\end{figure*}

\begin{figure}
    \centering
    \includegraphics[width=\linewidth]{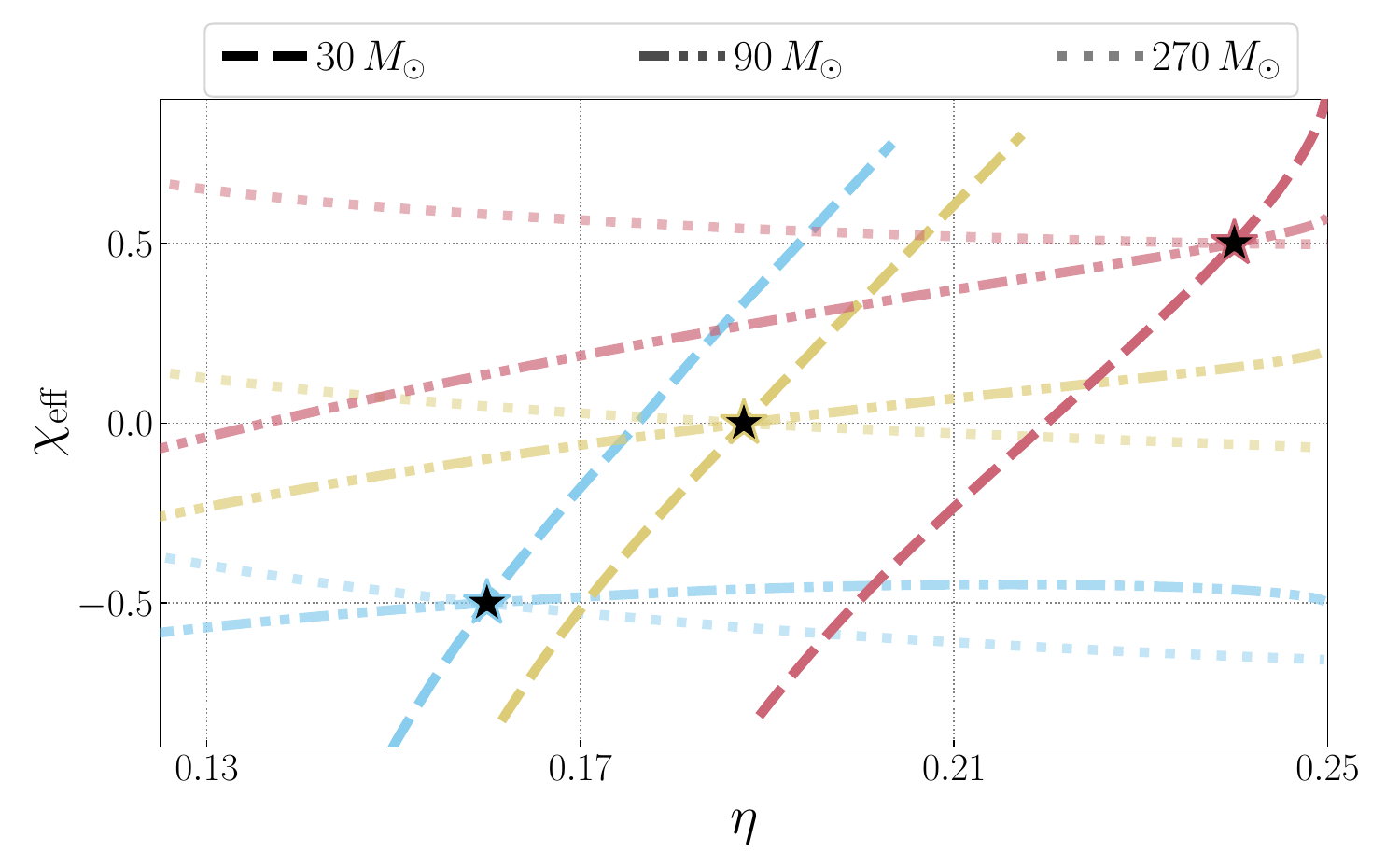}
    \caption{Comparison of the $\chieff-\eta$ correlation for signals with total mass $30\,M_{\odot}$, $90\,M_{\odot}$, $270\,M_{\odot}$ (darker to lighter shades of the same color for each reference point). We show results for the same three reference points as Fig.~\ref{fig:2D3D_30_chieffeta} but only include the 3D correlation paths computed from the aligned space of $\boldsymbol{x^{\rm(al)}}=[q,\chi_{1z},\chi_{2z}]$.
    The slope of the correlation dramatically decreases as the total mass increases.
    Fits to the correlation paths are provided in the text.}
     \label{fig:30_90_270_chieffeta}
\end{figure}

Having recovered spin-aligned correlations with the different networks, we now turn to cross-network comparisons.
Such comparisons reveal how correlations change as the observed signal becomes increasingly merger-dominated.
Figure~\ref{fig:3D_30_90_270_projected} shows correlation paths recovered in the 3D aligned spin space and then projected into the 2D subspaces formed from $[q,\chi_{1z},\chi_{2z}]$ for all three different total mass test cases.
Figure~\ref{fig:30_90_270_chieffeta} further projects down to the $[\eta, \chieff]$ space and demonstrates how the $\chieff{-}\eta$ correlation is altered as the signal's total mass increases.

For $90\,M_{\odot}$ and $30\,M_{\odot}$ signals, the slope of the correlation path in $\chieff-\eta$ is positive: a larger aligned spin is degenerate with more equal masses.
We interpret the sign of the correlation again through the effect of aligned spin and mass ratio on the signal length.
The slope of the correlation paths decreases with increasing total mass from $30\,M_{\odot}$ to $90\,M_{\odot}$.
For signals where more of the merger is observable, a change in the mass ratio can be ``counteracted" by a smaller change in the aligned spin.
Stated differently, at a fixed total mass, the aligned spin is better measured than the mass ratio for $90\,M_{\odot}$ signals, while the opposite is true for $30\,M_{\odot}$ ones.
This can be seen in Fig.~\ref{fig:30_90_270_chieffeta}: due to the correlations' ``steep slope," the paths from the $30\,\Msun$ signals span the full $\chieff$ regime of $-1$ to $1$, but only a subset of the allowed $\eta$ values; the inverse is true for the $90\,M_{\odot}$ case.

At $270\,M_{\odot}$, the correlation is mostly flat with a small \textit{negative} slope, i.e.,~larger aligned spins correspond to more unequal masses, the inverse of the correlation direction found for the lower masses cases. 
However, the trend is weak, suggesting that at a fixed total mass, the aligned spin is well-measured, while the mass ratio remains unmeasurable. 
We interpret this as follows: when the inspiral is no longer observable, $\chieff$ is no longer obtained from the inspiral length, but from the merger frequency. 
If the total mass, which also affects the merger frequency, is fixed, $\chieff$ can be measured accurately.
The mass ratio, on the other hand, has a subdominant effect on the merger frequency and is thus more challenging to measure without seeing the inspiral.

Finally, we fit the $\chieff{-}\eta$ correlation paths of Fig.~\ref{fig:30_90_270_chieffeta} with the ansatz 
\begin{align}
    \chieff&=(a_2M^2+a_1M+a_0)\eta^2\nonumber\\
&+(b_2M^2+b_1M+b_0)\eta\nonumber\\
&+(c_2M^2+c_1M+c_0)\,.
\label{eq:fit_chieffeta}
\end{align}
The fitting formula allows for a quadratic relation between $\chieff$ and $\eta$ where the coefficients are themselves quadratic functions of the total mass $M$.
We obtain 
$$
\begin{array}{lll}
    a_2=-5.6\times10^{-3}, & a_1=2.2, & a_0 =-1.7\times 10^2, \\
    b_2 =3.9\times10^{-3}, & b_1 =-1.5, & b_0=1.1\times 10^2, \\
    c_2=-4.9\times10^{-4}, & c_1=0.2, & c_0=-14\,,
\end{array}
$$
for the reference system $(q, \chi_{1z},\chi_{2z}) = (4,-0.5,-0.5)$.
Other systems can be approximately fit by shifting this fit vertically by an amount that depends on the new reference value.
Though this fit is based on a handful of reference systems and values of the total mass,\footnote{For this reason we also do not comment on its accuracy. As we are fitting three values of the total mass with three parameters, the fit's error or residual might be deceptive.} it represents a preliminary quantitative (albeit phenomenological) description of the imprint of spins on GWs across total masses.
Generalizing the fits to more dimensions and reference points is left for future work.

\section{Precessing degrees of freedom}
\label{sec:precession}

\begin{figure*}
    \centering
    \includegraphics[width=\linewidth]{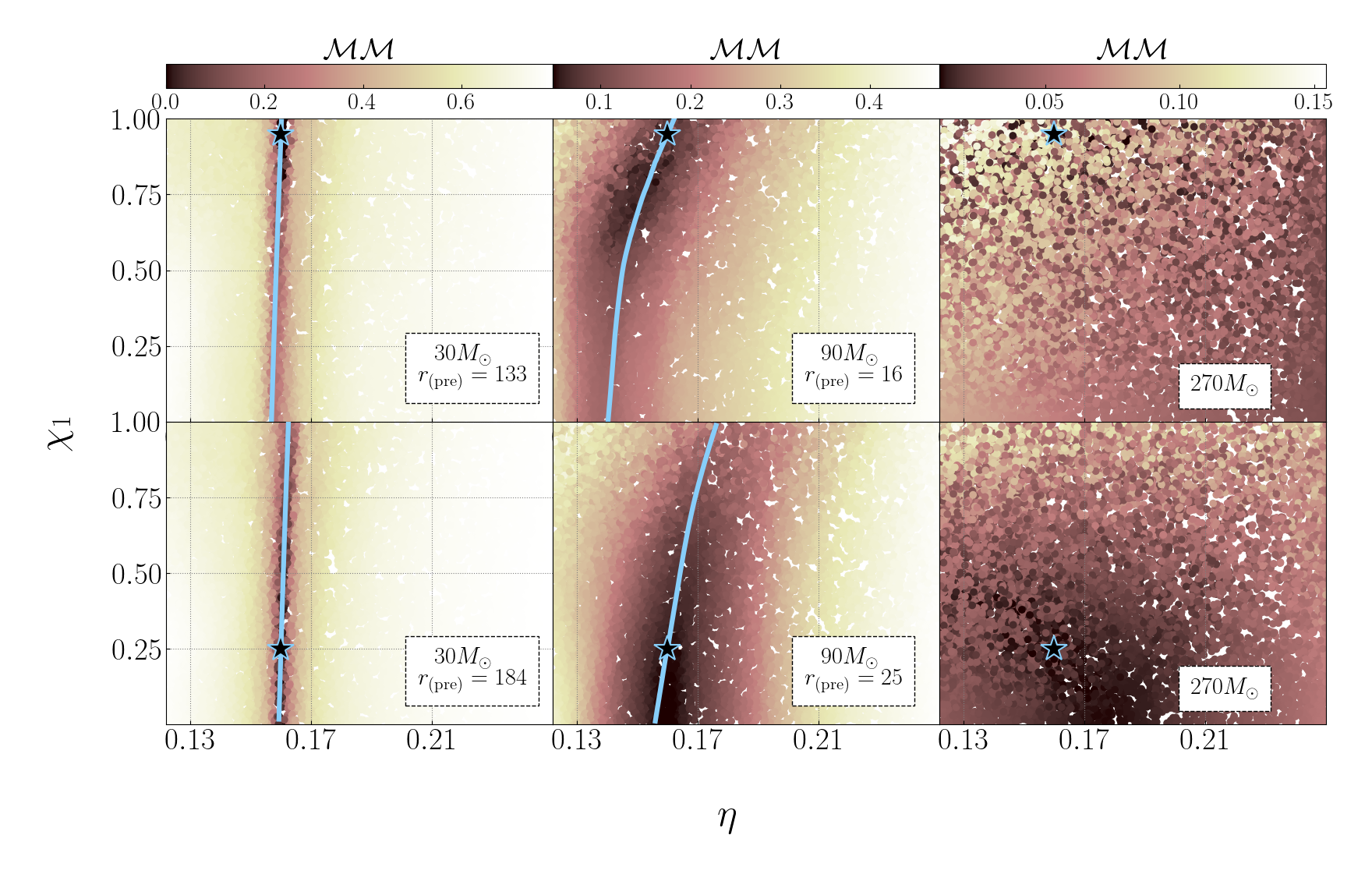}
    \caption{Comparison of precessing correlations for systems with a total mass of  $30\,M_{\odot}$ (left), $90\,M_{\odot}$ (middle), and $270\, M_\odot$ (right)  for a reference point with $\left(\eta, \chi_{1}\right)=(0.16,0.95)$ (top) and $(0.16,0.25)$ (bottom).
    The colormap denotes the mismatch (spanning a reduced range compared to previous results for readability) with respect to the reference point (star).
    Correlations are mapped in $\boldsymbol{x^{\rm (pre)}}=\left[\eta, \chi_{1}\right]$ with the secondary spin set to zero and the primary spin along the binary $x$-axis.
    Blue lines are the correlation path.
    The correlation strengths are given in-plot for the region of the reference point; the correlation strength varies along the correlation path. 
    No correlation path or strength is provided for the $270\,M_{\odot}$ case as no structure exists in the mismatches.
    Similar to the aligned-spin case, both the strength and the shape of the correlation between in-plane spins and mass ratio changes with the binary total mass.}
     \label{fig:2D_prec_30_90_270}
\end{figure*}

We now turn to the precessing parameter space, $\boldsymbol{x^{\rm (pre)}}$.
As a reminder, we here set $\vec{\chi}_2$ and $\phi_1$, in addition to the aligned components, to zero.
Under this restriction, the effective precessing parameter $\chip$ reduces to the primary spin magnitude $\chi_1$, which is the same as $\chi_{1x}$; see Eq.~\eqref{eq:chip}.
Figure~\ref{fig:2D_prec_30_90_270} compares precessing correlations for two unequal-mass reference systems across total masses: one with near-maximal $\chi_1$ (top row) and one with a smaller value (bottom row).
The mismatches (whose plotted colormap ranges have been restricted for readability) show clear structure for the $30\,M_{\odot}$ and $90\,M_{\odot}$ cases, but little structure for $270\,M_{\odot}$.
The latter is consistent with the fact that precession can only be measured for specific, fine-tuned configurations for merger-dominated signals~\cite{Biscoveanu:2021nvg}.\footnote{Additionally, over the plotted parameter space, mismatches vary by ${\sim}0.1$ which is within a factor of $10$ of the expected network error, listed in Table~\ref{table:mae_models}.} 
As such, for the rest of the section we only consider $30\,M_{\odot}$ and $90\,M_{\odot}$ binaries.

We map correlations in the 2D precessing space of $\boldsymbol{x^{\rm (pre)}}=\left[\eta, \chi_{1}\right]$.
For $30\,M_{\odot}$, the correlation strength is $r_{\rm(pre)}={\cal{O}}(100)$.\footnote{The precessing correlation strength $r_{\rm(pre)}$ is not comparable to the effective aligned correlation $r_{\rm(eff)}$ presented in Sec.~\ref{sec:inspiral}. Even though both refer to correlations in the same number of dimensions, they vary or set to zero different parameters.}
Given that the correlation path is almost vertical in the $\chi_1{-}\eta$ plane, this large $r$ value reflects that the mass ratio can be measured accurately.
The slight tilt of the $30\,\Msun$ correlation path indicates, however, that there is some degree of correlation between the in-plane spin and the mass ratio: either drastically decreasing the in-plane spin magnitude or making the masses slightly more unequal both alter the waveform in similar ways.
The $90\,M_{\odot}$ correlation path tilts further from the vertical and the correlation strength reduces to $r_{\rm(pre)}={\cal{O}}(10)$. 
Moreover, the correlation strength changes along the correlation path, becoming weaker (lower mismatches on average) away from the reference point.

\begin{figure}
    \centering
    \includegraphics[width=\linewidth]{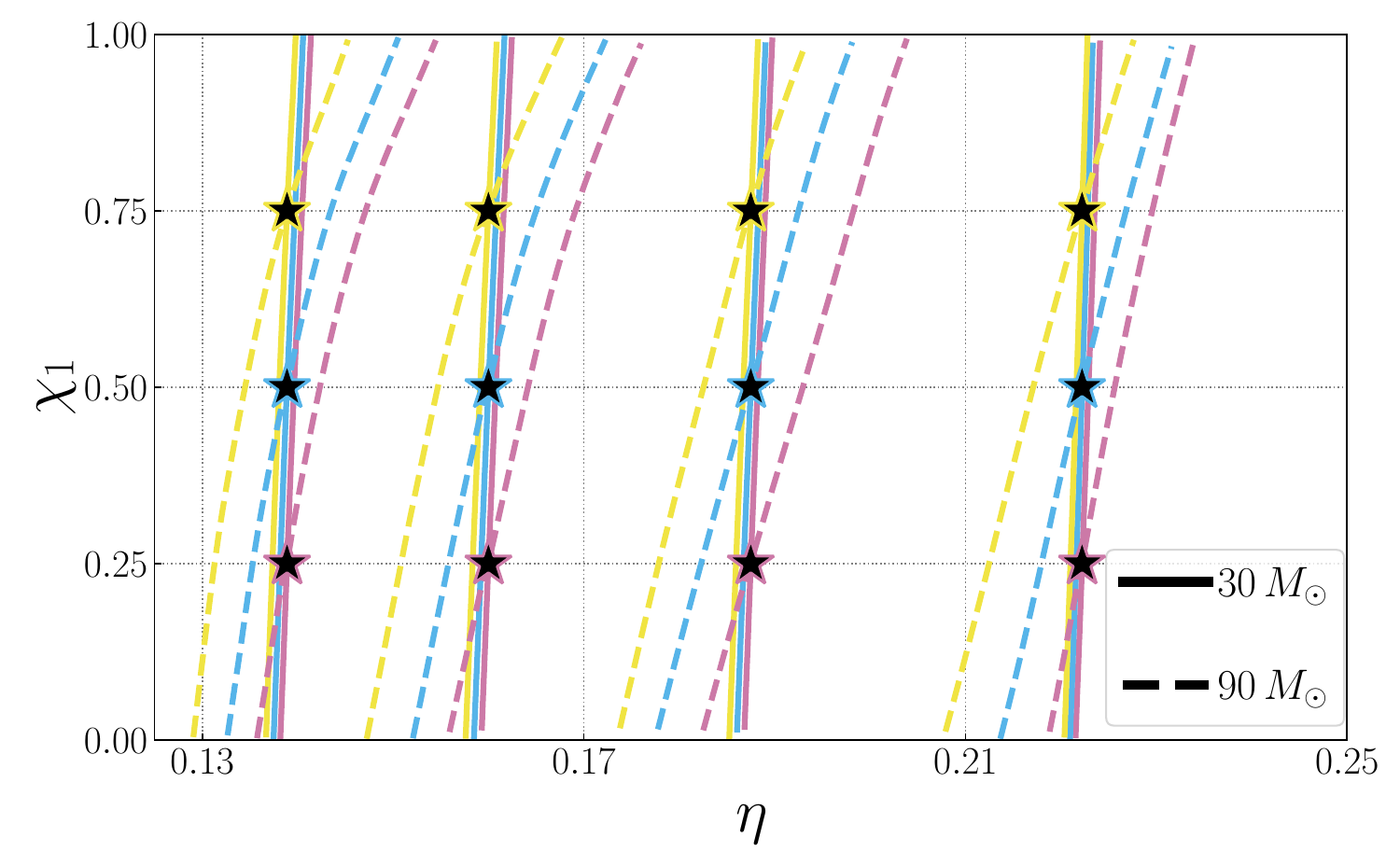}
    \caption{Comparison of the precessing correlation for signals with total mass $30\,M_{\odot}$ (solid) and $90\,M_{\odot}$ (dashed). We show results for various reference points (stars) and correlations mapped in $\boldsymbol{x^{\rm (pre)}}=\left[\eta, \chi_{1}\right]$.  Colors indicate the value of $\chi_1$ at reference, with pink, blue, and yellow representing $\chi_1 = 0.25, 0.5, 0.75$ respectively. 
    From left to right, we consider $\eta =0.14, 0.16, 0.19, 0.22$ (equivalently, $q=5,4,3,2$). 
    The change in correlation shape shown in Fig.~\ref{fig:2D_prec_30_90_270} is a generic feature across multiple reference points.}
     \label{fig:2D_prec_all}
\end{figure}

Similar to the aligned-spin results of Secs.~\ref{sec:inspiral} and~\ref{sec:merger}, the correlation between the in-plane spin and the mass ratio both changes in shape and reduces in strength as the total mass increases.
Put differently, as the total mass increases, the impact of mass ratio on the waveform morphology diminishes, making $\eta$ harder to measure.
Since $\chi_1$ is comparably constrained at $30\,\Msun$ and $90\,\Msun$, this increased uncertainty on $\eta$ makes the over-all correlation \textit{strength} weaker.
However it also reveals more correlational \textit{structure} in the $\chi_1{-}\eta$ plane: a smaller change in $\chi_1$ is now sufficient to mimic the same variation in $\eta$.
The shape of the correlation across more reference points is explored in Fig.~\ref{fig:2D_prec_all}.
We plot correlation paths for different reference points (stars and different colors) and values of the total mass (solid vs dashed lines).\footnote{Since we only have results for two values of the total mass, we do not present fits of the paths of Fig.~\ref{fig:2D_prec_all}, leaving them to future extensions.}
Consistent, with Fig.~\ref{fig:2D_prec_30_90_270}, the correlation paths for inspiral-dominated $30\,M_{\odot}$ systems are mostly vertical, while the tilt of the path increases at $90\,M_{\odot}$ for all reference points.
Our results suggest that when less inspiral is observable, an appropriate combination of $\eta$ and $\chi_1$ should be better measured than either parameter alone.

\begin{figure}
    \centering
    \includegraphics[width=\columnwidth]{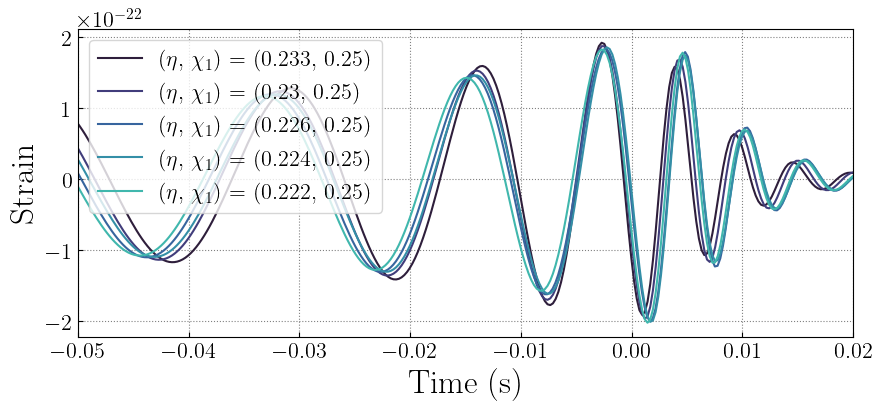}
    \includegraphics[width=\columnwidth]{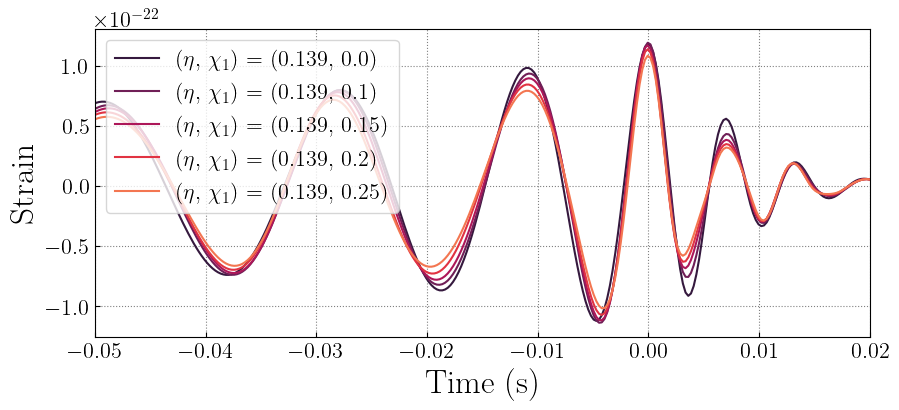}
    \caption{A demonstrative example of how small variations in $\eta$ and in-plane $\chi_1$ effect signal morphology in the late inspiral and merger. We show sets of waveforms with fixed $\chi_1=0.25$ and varied $\eta$ (top), compared to fixed $\eta=0.139$ ($q=5$) and varied $\chi_1$ (bottom).
    All waveforms are generated using \textsc{NRSur7dq4} with a total mass of $90\,\Msun$ and all spin degrees of freedom set to 0 aside from the primary's in-plane spin along the $x$-axis.}
    \label{fig:precessing_waveforms}
\end{figure}

We explore the underlying cause of these{correlations} by looking directly at waveform morphology in the late inspiral and merger. 
Figure~\ref{fig:precessing_waveforms} shows $90\,M_{\odot}$ waveforms with fixed versus varied $\eta$ and $\chi_1$. 
Empirically, both $\eta$ and $\chi_1$ subtly change the waveform phase evolution and alter the structure of the ringdown. 
At a fixed $\chi_1$, increasing $\eta$ (making the masses more equal; lighter to darker colors) slightly \textit{increases} the merger frequency (top panel). 
To identify this, draw your eye to the peak cycle where waveforms are aligned (between -0.01 and 0\,s), and compare these to the location of the preceding peaks/troughs. 
A lower merger frequency corresponds to more time between these local extrema.
At a fixed mass ratio, increasing $\chi_1$ (darker to lighter colors) slightly \textit{decreases} the merger frequency (bottom panel). 
This effect is stronger for more unequal masses.
Combining these observations suggests that increasing both parameters along the line of correlation in the $\chi_1{-}\eta$ plane leaves the merger frequency essentially unchanged.
In agreement with Fig.~\ref{fig:2D_prec_all}, a small change in $\eta$ affects the waveform much more than a corresponding change in $\chi_1$.
Given that precession-driven amplitude fluctuations~\cite{Kidder:1993} are often unobservable in short, merger-dominated signals,
the merger frequency shift could become a primary precession observable.
The question of how exactly precession is measured in the merger remains open, but observations like this offer a qualitative understanding of the link between parameter correlations and waveform morphology.
Correlation mapping is a systematic approach to identifying and understanding these relations.

\section{Conclusions}\label{sec:conclusions}

In this study, we proposed a waveform-based approach to study the imprint of spin degrees of freedom on GWs emitted by BBHs.
Our approach does not require analytic expressions for the binary dynamics, it is instead based on the availability of forward models for the emitted signal given the binary parameters.
It is therefore applicable even to the merger for which solutions can only be obtained via full numerical relativity simulations.
The method is based on identifying parameter correlations, i.e., paths in parameter space that leave the signal approximately unaltered. 

{Rather than focusing on \textit{a single point in the parameter space}, as in traditional parameter estimation, our framework traces \textit{correlation paths} across a deliberately restricted subset of the intrinsic parameter space.
This simplification provides a systematic way to map spin correlations across broad regions of parameter-space and study merger dynamics without relying on analytic models.} Studying the spin aligned and precessing degrees of freedom for inspiral-dominated (total mass $30\,M_{\odot}$), late-inspiral-plus-merger ($90\,M_{\odot}$), and merger-dominated ($270\,M_{\odot}$) signals, c.f. Fig.~\ref{fig:waveforms}, we find the following.
\begin{itemize}
    \item Mapping correlations in the generic 3D aligned space of $\boldsymbol{x^{\rm(al)}}=[q,\chi_{1z},\chi_{2z}]$ can reveal the known lower-dimensional correlations of the 2D effective space of $\boldsymbol{x^{\rm(eff)}}=[\eta, \chieff]$ for inspiral-dominated signals; see Fig.~\ref{fig:2D3D_30_chieffeta}.
    \item The effective aligned spin $\chieff$ captures the aligned spin dynamics for systems up to ${\sim}90\,M_{\odot}$. 
    However, at $90\,M_{\odot}$, the aligned spin-mass ratio correlation is comparable or stronger in the 3D aligned space of $\boldsymbol{x^{\rm(al)}}=[q,\chi_{1z},\chi_{2z}]$ than the 2D effective space of $\boldsymbol{x^{\rm(eff)}}=[\chieff,\eta]$; see Fig.~\ref{fig:2D3D_90_chieffeta}. This suggests that a more appropriate parameter might exist.
    \item For systems with a total mass of $270\,M_{\odot}$, correlation mapping using the full aligned spin degrees of freedom does not yield the same results as working directly with $\chieff$; see Fig.~\ref{fig:2D3D_270_chieffeta}. This strongly suggests that $\chieff$ is a suboptimal aligned effective spin for merger-dominated signals~\cite{Healy:2018swt,Pratten:2020fqn}.
    \item As the binary total mass increases, the correlation between the the aligned spins and the mass ratio becomes less steep, see Figs.~\ref{fig:3D_30_90_270_projected} and~\ref{fig:30_90_270_chieffeta}.
    The corresponding fit of Eq.~\eqref{eq:fit_chieffeta} is a phenomenological description of aligned spin dynamics that is applicable even for systems with a (detector frame) total mass of $270\,M_{\odot}$ and for which only the merger is observable.
    \item Similarly to the aligned-spin case, the correlation between the in-plane spin and the mass ratio becomes less steep and reduces in strength as the total mass increases.
    The correlation has effectively disappeared by a total mass of $270\,M_{\odot}$, supporting known difficulties of measuring spin precession in merger-dominated signals~\cite{Biscoveanu:2021nvg,Xu:2022zza,Vitale:2014mka,Shaik:2019dym,Gerosa:2015tea}.
\end{itemize}

The correlation mapping procedure can be extended along various directions.
Firstly, though we explore results at different values of the total mass, its value remains fixed within each mapping.
This restriction ignores correlations between the total mass and other parameters.
Preliminary results with a neural network that includes the total mass are promising in that they can reproduce the spin-mass ratio correlation known from parameter estimation, see Sec.~\ref{sec:inspiral}. 
However, technical challenges remain related to the boundaries in the mass parameter space where we need to switch between the two waveform models considered, \textsc{IMRPhenomXPHM} and \textsc{NRSur7dq4}.
Secondly, with the total mass varying, it would be preferable to switch to defining the spin directions at a fixed point in the waveform, e.g., $t=-100\,M$ rather than a fixed frequency~\cite{Varma:2021csh}.
Thirdly, in this study we only consider the intrinsic degrees of freedom. 
The measurability of spins also depends on extrinsic binary parameters~\cite{Fairhurst:2019vut}; for example, $\chip$ and the binary's orientation in the sky have known observational degeneracies, e.g.,~\cite{Fairhurst:2023idl,Miller:2025eak}, making correlations between spins and extrinsic degrees of freedom interesting to consider{, particularly the aforementioned effects of inclination-dependent mode mixing in precessing systems.} \
Fourthly, we are looking for correlation \emph{paths}. 
As we look for correlations in higher dimensions, these should be correspondingly generalized to higher-dimensional correlations, e.g., correlation \textit{surfaces}.
Finally, Eq.~\eqref{eq:fit_chieffeta} provides a first quantitative result on the imprint of spins applicable even to merger-dominated signals, but it is still restricted to $\chieff$ and $\eta$.
Higher dimensional fits of the 3D aligned spin parameters or the precessing parameters could motivate better effective parameters.
{In particular, we aim to derive better-measured spin parameters for high-mass signals--especially to capture the precessing degrees of freedom. }

The results of Figs.~\ref{fig:2D3D_90_chieffeta} and~\ref{fig:2D3D_270_chieffeta} indeed suggest that a different effective aligned spin parameter than $\chieff$ might be more appropriate for merger-dominated signals. 
Studies of the ringdown and final-state of spin-aligned systems show that post-merger properties and phenomenology can be well-captured by functions of the pre-merger masses and spins~\cite{Healy:2018swt, Cheung:2023vki, Jimenez-Forteza:2016oae}. 
This implies that combinations of mass ratio and aligned spin that lead to the same \textit{remnant} spin may, at a fixed total mass, be a better measured parameter combination for merger/ringdown dominated signals.
Additionally, studies of the final-state of waveforms in the \textsc{IMRPhenom} family show that the late inspiral and merger are better characterized by an effective \textit{total} spin parameter~\cite{Pratten:2020fqn}. 
Even in the inspiral phase, there are effective-aligned spin parameters that may capture dynamics better than $\chieff$; e.g.,~a normalized version of the exact 1.5\,PN coefficient, which directly contributes to the signals' phase evolution~\cite{Pratten:2020fqn,Ossokine:2017dge}.
We leave exploring these alternate parametrizations, and extending beyond them, to future work.

{
Insights from correlations paths can aid in waveform modeling through identification of further effective spin parameters like $\chieff$, $\chip$, and others~\cite{Pratten:2020fqn,Ossokine:2017dge,Thomas:2020uqj,Gerosa:2020aiw}.
Using fits like Eq.~\eqref{eq:fit_chieffeta} (after extending to more values for the total mass) can lead to phenomenological definition of effective parameters that are appropriate for heavy BBHs.
Such effective parameters can improve the ansatz for waveforms calibration to numerical relativity simulations~\cite{Pratten:2020ceb} or help simplify the dynamics~\cite{Hannam:2013oca}.
Additionally, better parameters to describe the late inspiral and merger regime can lead to more accurate predictions for remnant BH properties~\cite{Thomas:2020uqj}.
Finally, while we have focused on correlation paths via the covariance eigendirection with the largest eigenvalue, the full matrix contains information about the parameter structure.
Specifically, studying the orthogonal directions along the correlation path would indicate where the mismatch grows most rapidly and thus point to 
dominant modeling effects.

{Correlation paths can also help interpret and improve parameter estimation results. 
First, Fig.~\ref{fig:30_90_270_chieffeta} shows how the $\chieff$-$\eta$ degeneracy changes at high mass. 
This is echoed in parameter estimation results, c.f., Fig.~5 of \citet{GWTC1} where low-mass binaries have a steeper slope in the $\chieff$-$q$ plane than the high-mass ones.
Such ``consistency with expectations'' tests can help us safeguard against biases: a posterior that does \textit{not} follow these known paths  could indicate that the signal may be contaminated by non-Gaussian noise~\cite{Ashton:2021tvz} or that waveform systematics are at play.} 
{Second, $\chieff$ and $\chip$ are not just beneficial for waveform modeling, but are also used in parameter estimation--for both individual BBHs and their population at large. 
The tilt angles, for example, of BBHs are poorly constrained on a population level~\cite{Vitale:2022dpa,Miller:2024sui}, but the $\chieff$ distribution--also used to probe anti-aligned spin--is much more precisely constrained~\cite{KAGRA:2021duu}.}
{Lastly, as shown in Fig.~\ref{fig:2D3D_30_chieffeta}, the mismatch still varies along the correlation path, indicating that these are only approximate degeneracies. 
A sufficiently high SNR will eventually resolve them. Though not analyzed here, our framework provides a structure to quantify this SNR. }

Measurements of spins from massive systems have inspired exciting astrophysical interpretations.
Understanding how precession is imprinted on these signals and what parameter combinations drive the signal morphology will remain important as more events are detected.
Waveform-based correlation mapping provides a principled way to motivate the form of aligned- and precessing-spin effective parameters across masses.

\section*{Data Availability}

The $30, 90, 270\, \Msun$ neural networks used to generate all our results are found at Ref.~\cite{github_release}. This {\tt github} repository also includes notebooks and scripts to run our correlation mapping algorithm and visualize results.

\section*{Acknowledgements}

We thank Nicole Khusid, Max Isi, and Geraint Pratten for helpful comments and suggestions, as well as Lucy Thomas{,} Aaron Johnson{, and Rhiannon Udall} for their collaboration and insights about waveform modeling and glitches.
This work was supported by NSF Grant PHY-2150027 as part of the LIGO Caltech REU Program which funded KK.
SM and KC were supported by NSF Grant PHY-2308770 and PHY-2409001.
DF was supported by NSF Grants OAC-2004879, PHY-2207780, and PHY-2114581.
The authors are grateful for computational resources provided by the LIGO Laboratory and supported by National Science Foundation Grants PHY-0757058 and PHY-0823459.
This material is based upon work supported by NSF's LIGO Laboratory which is a major facility fully funded by the National Science Foundation.
Software: \texttt{LALsuite}~\cite{lalsuite}, \texttt{matplotlib}~\cite{Hunter:2007}, \texttt{sklearn}~\cite{scikit-learn}, \texttt{Tensorflow}~\cite{tensorflow2015-whitepaper}, \texttt{numpy}~\cite{harris2020array}, \texttt{pandas}~\cite{reback2020pandas}, \texttt{sympy}~\cite{10.7717/peerj-cs.103}.

\appendix

\section{Neural network construction}
\label{app:network}

\begin{table}[]
\centering
\begin{tabular}{c|ccccc}
\toprule
Mass ($M_\odot$) & 30 & 90  & 270 \\
\midrule
Waveform & \textsc{IMRPhenomXPHM} & \multicolumn{2}{c}{\textsc{NRSur7dq4}} \\

\midrule
Train & 0.00728 & 0.0103 & 0.00577 \\
Development & 0.00862 & 0.0139  &  0.00900 \\
Test & 0.00954 & 0.0139  & 0.00939 \\
\bottomrule
\end{tabular}
\caption{Performance of the three neural networks used in this work. Each network predicts the mismatch between two signals with the same fixed total mass (either $30$, $90$, or $270\,\Msun$) but different mass ratios and spins. We list each network's total mass, relevant waveform model, and the mean absolute error for the mismatch from each set.}
\label{table:mae_models}
\end{table}

The mismatch neural networks are constructed following the network of~\citet{Ferguson:2022qkz}, extended to include higher order modes, more values for the total mass, and parametrized waveform models rather than NR simulations.
Figure 1 of~\citet{Ferguson:2022qkz} and the relevant discussion therein provide details on the network settings that we adopt here.

\begin{figure}[]
    \centering
    \includegraphics[width=\linewidth]{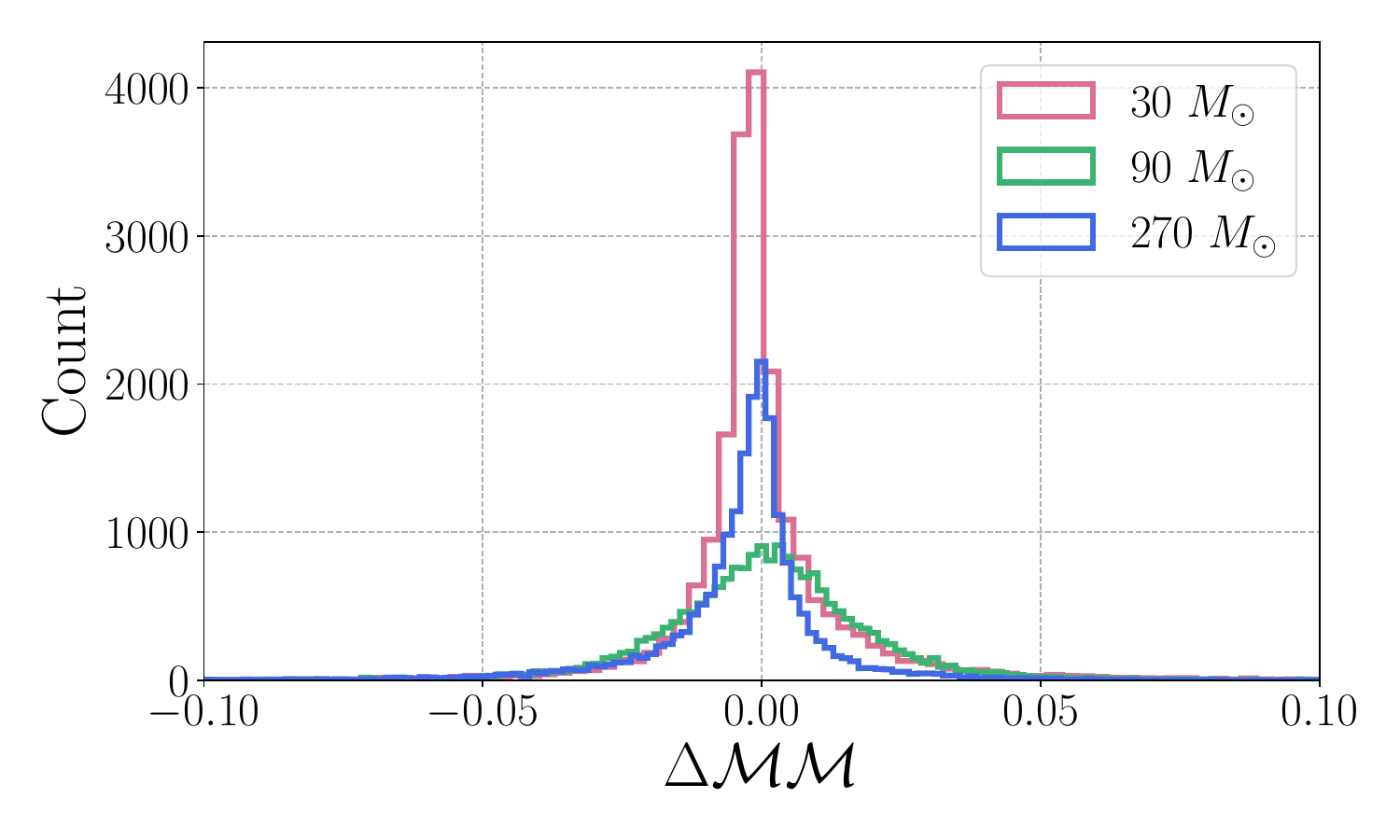}
        \caption{Distribution of the difference between the exact mismatch and the predicted mismatch, $\Delta {\cal{MM}}$,  for waveforms of the test set for the three networks with different total masses. The $3$ networks perform similarly with standard deviations $0.0181, 0.0196$, and $0.0152$ for total masses of $30\,M_{\odot}, 90\,M_{\odot}$, and $270\,M_{\odot}$, respectively.} 
    \label{fig:MMverification}
\end{figure}

For each detector-frame total mass value, we generate $2000$ samples from the parameter space uniformly in $\eta\in[0.12,0.25]$ (corresponding to $q\leq6$), $\chi_i\in[0,1)$, and on the unit sphere for spin directions.
We form system pairs without repetitions and split them into training, development, and testing sets with relative ratios of $8:1:1$. 
For each pair, we compute the mismatch with the appropriate waveform model including higher-order modes using Eq.~\eqref{eq:mismatch} with $f_0=20\,$Hz, a flat noise spectrum, and a face-on orientation while maximizing over time and phase. 
The spin reference frequency is $f_{\mathrm{ref}} = f_0$. 
Table~\ref{table:mae_models} summarizes the properties of each network: its total mass, waveform model it was constructed with, and the mean absolute error for the mismatch in each set.
\begin{figure*}
    \centering
        \includegraphics[width=\linewidth]{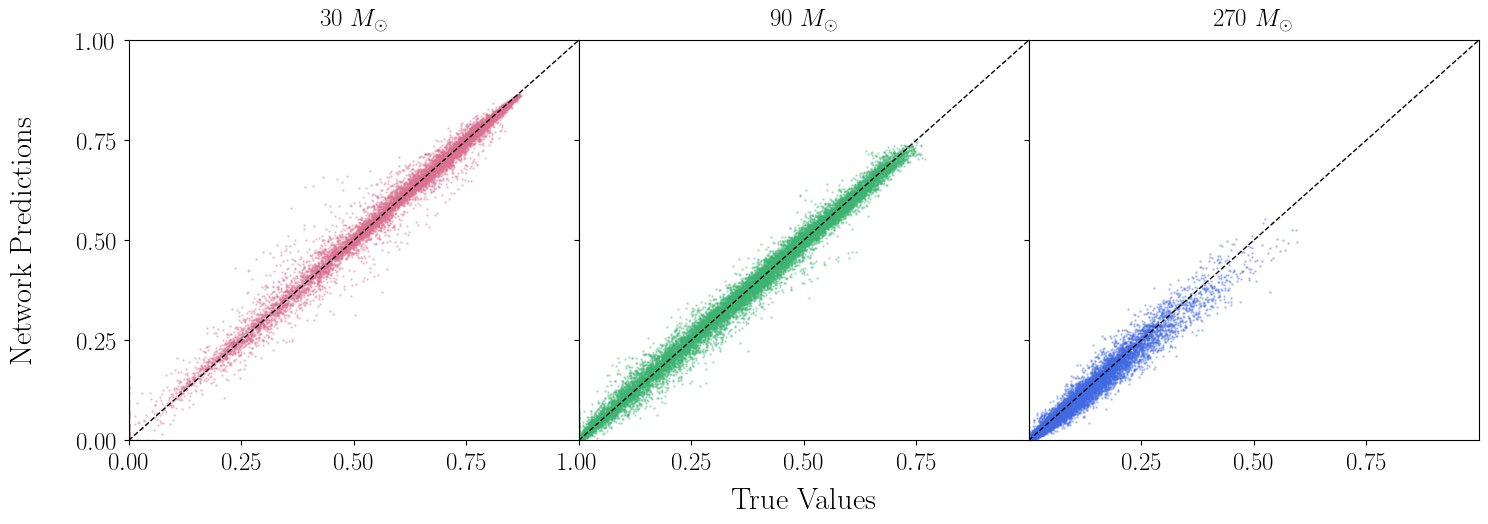}        
        \caption{Scatter plot of network-predicted mismatches with true mismatches computed directly from the waveforms in the test set using Eq.~\eqref{eq:mismatch}. The diagonal black dashed line represents the ideal case where predicted and true mismatch values are equal. Results are shown for the three networks corresponding to systems with total masses of 30\,$M_\odot$ (left), 90\,$M_\odot$ (middle), and 270\,$M_\odot$ (right).}
    \label{fig:MMpredictions}
\end{figure*}

\begin{figure*}
    \centering
        \includegraphics[width=\linewidth]{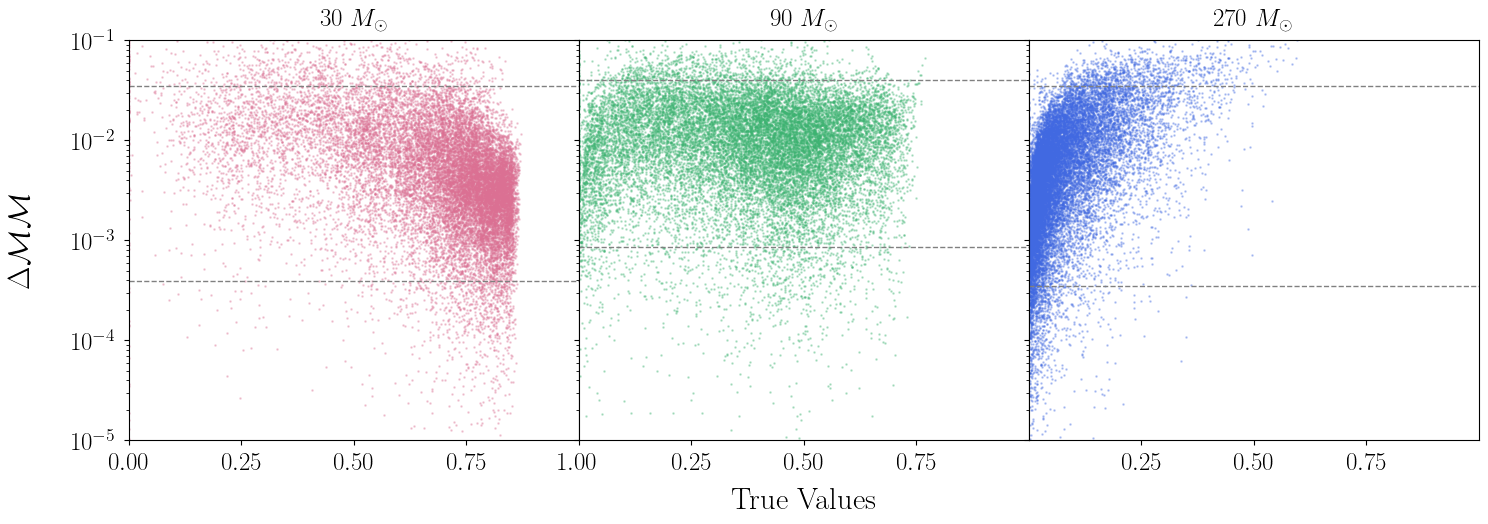}        
        \caption{{Scatter plot of the absolute error in mismatch predictions with respect to the true mismatch values, corresponding to the same test set shown in Fig.~\ref{fig:MMpredictions}. The horizontal dashed lines indicate the 5th and 95th percentiles of all points along the $y$-axis, marking the central 90\% interval of absolute error.}}
    \label{fig:MMerror}
\end{figure*}

The distribution of the difference between the exact mismatch from Eq.~\eqref{eq:mismatch} and the network prediction is shown in Fig.~\ref{fig:MMverification}. 
The standard deviations of the distributions are $0.0181, 0.0196$, and $0.0152$ for the networks with total masses of $30\,M_{\odot}, 90\,M_{\odot}$, and $270\,M_{\odot}$, respectively. 
All networks perform similarly, with the $270\,M_{\odot}$ case (merger-dominated signals) having the smallest mismatches altogether. 
This is likely due to the short duration of the signal, making it less sensitive to parameter variations. 
A scatter plot of the true mismatch versus the network prediction is shown in Fig.~\ref{fig:MMpredictions}. 
The highest mass network, for signals dominated by the merger phase, exhibits the smallest maximum mismatch, below ${\sim}0.7$. 
In contrast, lower-mass systems, which are inspiral-dominated, have a wider spread of signal morphology and therefore mismatch. 
The difference between the true and the predicted value is larger for medium mismatch values, around 0.5. Large (very different waveforms) or small (very similar waveforms) mismatches are easier to predict. {In Fig.~\ref{fig:MMerror}, we plot the absolute error with respect to the true mismatch values and indicate the 90\% confidence interval. Across all total masses, the absolute errors are at the percent level or lower.}

\section{Fitting the mismatch distribution}
\label{app:PCAGMM}
\begin{figure*}
    \includegraphics[width =\textwidth]{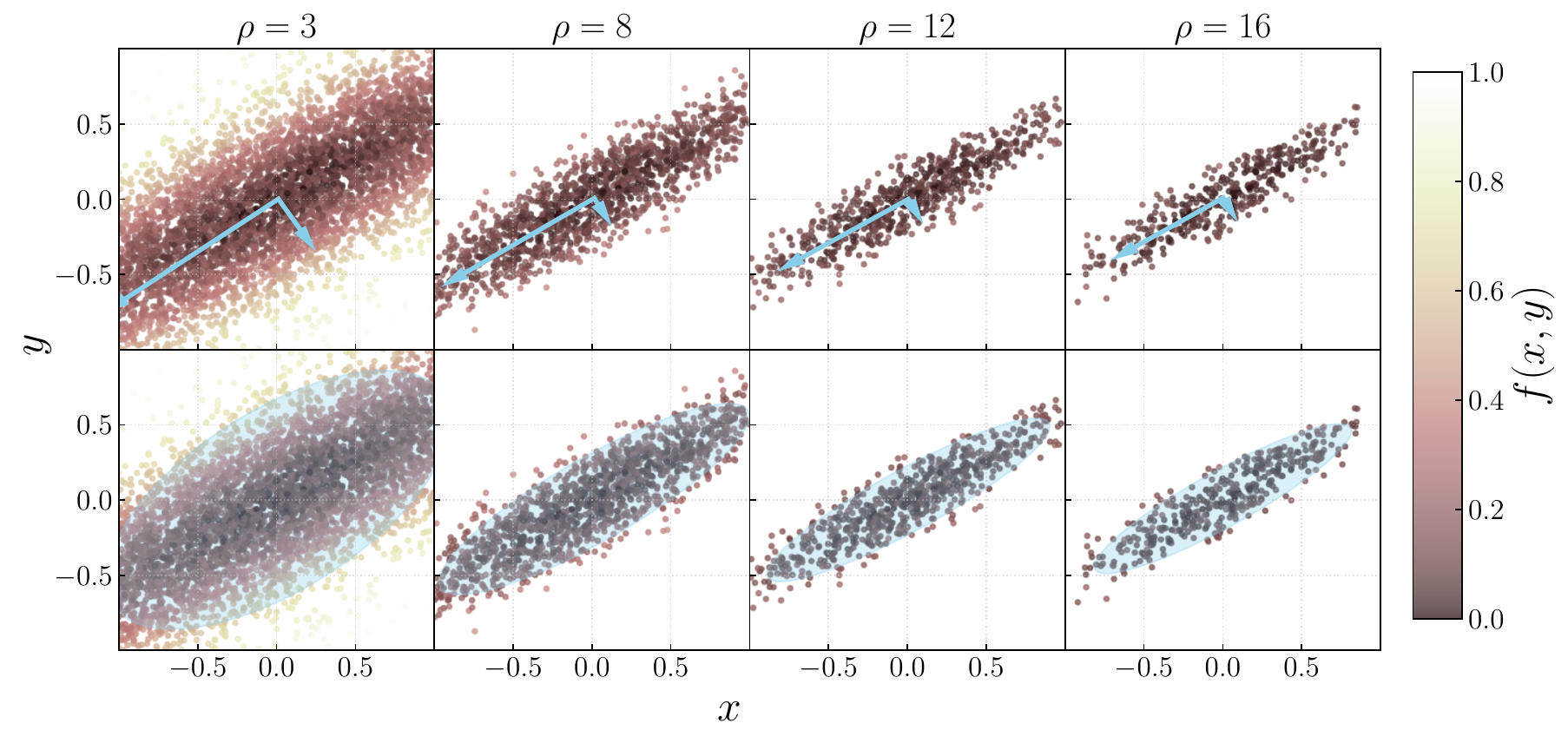}
    \caption{Toy model demonstrating the PCA (top) and Gaussian fit (bottom).
    Data are generated uniformly in the parameter space and then assigned a scalar value from a chosen Gaussian plus additive noise.
    We then apply the method of Fig.~\ref{fig:flowchart} and plot the principal components (top) and Gaussian covariance (bottom, shaded region indicates 1 $\sigma$) for different values of the rejection sampling SNR (left to right).
    The scattered dots illustrate the data that are fit after rejection sampling and are colored by their value.
    Both methods qualitatively recover the expected directions.}
     \label{fig:toymodel}
\end{figure*}

\begin{figure}
    \centering
    \includegraphics[width=\linewidth]{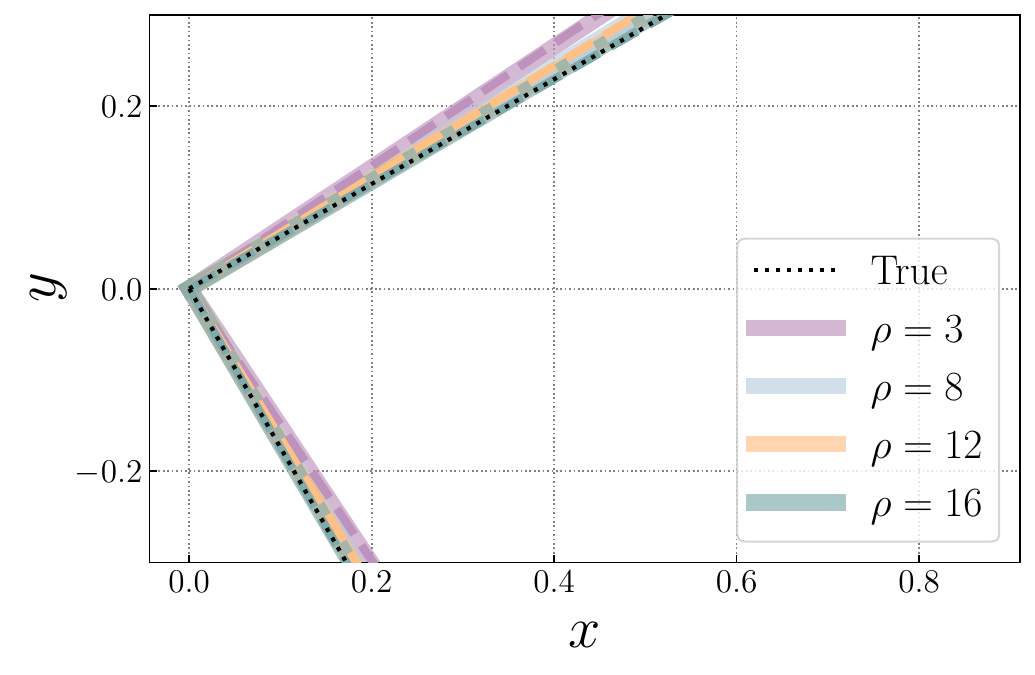}
    \caption{Comparison of the {correlation} tangent vectors derived from PCA (solid) and the Gaussian fitted covariance eigen vectors (dashed). 
    Different colors represent results for varying rejection sampling SNRs. 
    Black dotted lines indicate the true eigenvectors of the covariance matrix with which the data were simulated.
    PCA and Gaussian fitting return identical results and they further agree with the true directions for an appropriate SNR value.}
    \label{fig:components}
\end{figure}

We treat the mismatch distribution, e.g.~Fig.~\ref{fig:flowchart}, as a scalar function $f(\boldsymbol{x})$ where $\boldsymbol{x}$ denotes the source parameters. 
The mismatch distribution quantifies the degree to which different parameters lead to different waveforms with respect to the reference point $\boldsymbol{x}_0$.
The distribution peaks at the reference point and level curves represent parameter space regions where the mismatch remains constant, $\nabla f = 0$.
The tangent to the level curve at the reference point $\boldsymbol{v}$ is 
\begin{equation}
    \nabla f(\boldsymbol{x_0}) \cdot \boldsymbol{v}=0\,,
\end{equation}
and we define its direction as the{correlation} path in parameter space.
We therefore identify 1D{correlation} \emph{paths} rather than more extended multi-dimensional \emph{regions}.
Such regions would, for example, arise if some parameters (or some parameter combinations) have a negligible effect on the signal.

To extract these degenerate directions, we explore two methods: 1) Principal Component Analysis (PCA) and 2) Gaussian fit, both implemented through \texttt{sklearn}~\cite{scikit-learn}.
The main text results are based on Gaussian fitting.
In this Appendix, we compare the two methods' performance on a toy model.
From the PCA method, we adopt the first principal component, calculated using \texttt{sklearn.decomposition.PCA}, while from the Gaussian fit we use the covariance eigenvectors and eigenvalues using \texttt{sklearn.mixture.BayesianGaussianMixture}.

We generate synthetic data based on a 2D Gaussian distribution and random noise. 
The data are uniformly sampled in the range $[-1,1]$ for both the $x$ and $y$-dimensions. 
At each point $(x, y)$, we define a scalar function $f(x, y)$ using a multivariate Gaussian distribution with a known mean vector and covariance matrix.
We add noise uniformly sampled from the range $[-1,1]$ and scaled by a noise factor $\mu = 0.1$. 
The noisy scalar function is normalized to $[0,1]$, similar to the mismatch range.
Figure~\ref{fig:toymodel} shows the principal components (top) and Gaussian covariance (bottom) for different values of the rejection sampling SNR ($\rho$), Eq.~\eqref{eq:weights}. 
Both methods identify consistent directions given the same data. 
The tangent vectors from the PCA and eigenvectors of Gaussian covariance matrix are plotted in Fig.~\ref{fig:components}. 
For this toy model, rejection sampling with $\rho = 16$ most accurately recovers the true structure of the underlying distribution.
We repeat this exercise with toy models (different true distributions) and consistently find equivalent performance between the PCA and Gaussian fit methods.
\section{Correlations in the 3D aligned spin space}
\label{app:3Dsurface}
\begin{figure*}
\includegraphics[width=\linewidth]{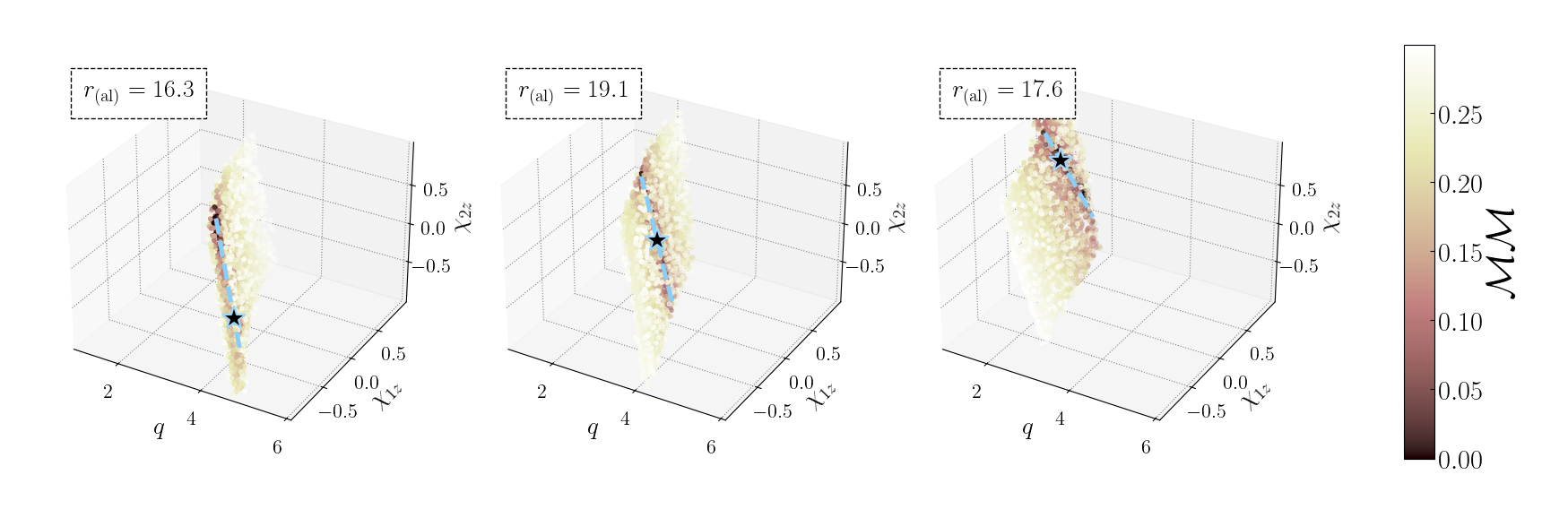}
    \caption{Similar to Fig.~\ref{fig:3D_30_qchi1zchi2z} but restricting to fitting the lowest-mismatch points. 
    We restrict to mismatches below $0.3$ to make the 3-dimensional structure more visible.
    Here, the recovered correlation path more closely aligns with the darker points, corresponding to  $[q,\chi_{1z},\chi_{2z}]$ points that result in $\chieff,\eta~{\sim}~$constant.
    }
\label{fig:3D_30_qchi1zchi2z_highSNR}
\end{figure*}
Low mismatches in the 3D aligned spin space of $[q,\chi_{1z},\chi_{2z}]$ are achieved in two ways.
The first group of low mismatch points consists of $[q,\chi_{1z},\chi_{2z}]$ such that $\chieff$ and $\eta$ remain constant. For example, keeping $q$ constant and varying the spins such that $(q\chi_{1z}+\chi_{2z})$ is kept nearly constant will result in systems with identical $\eta$ and $\chieff$ and thus the smallest mismatches.
These systems are the darkest dots in Fig.~\ref{fig:3D_30_qchi1zchi2z}.
The correlation mapping algorithm can fit those points if we increase the SNR of the rejection sampling step, i.e., restrict the fitting only to the lowest mismatch points. 
This is shown in Fig.~\ref{fig:3D_30_qchi1zchi2z_highSNR}.
Since these points keep $\chieff,\eta \sim $~constant, they do not trace the $\chieff{-}\eta$ correlation of Fig.~\ref{fig:2D3D_30_chieffeta}, instead staying at the reference value.

The second group of low-mismatch points includes $[q,\chi_{1z},\chi_{2z}]$ that vary in such a way that their $\chieff$ and $\eta$ follow the expected correlation of Fig.~\ref{fig:2D3D_30_chieffeta}.
Since these are more numerous than the first group (constant $\chieff$ and $\eta$), the Gaussian fit will preferentially identify them unless restricted to very low mismatches, e.g., Fig.~\ref{fig:3D_30_qchi1zchi2z_highSNR}.
These are the points presented in the main text in Fig.~\ref{fig:3D_30_qchi1zchi2z}, while Fig.~\ref{fig:2D3D_30_chieffeta} confirms that they indeed recover the expected $\chieff-\eta$ correlation.
Both groups of points could be fitted simultaneously by generalizing the correlation mapping to find surfaces rather than paths, which we leave to future work.

\bibliography{References}
\end{document}